\documentclass[prd,preprint,showpacs,eqsecnum,floatfix]{revtex4}

\usepackage{amsfonts}
\usepackage{amssymb}
\usepackage{graphicx}
\usepackage{pstricks}

\newcommand{\bpi}{\begin{picture}}
\newcommand{\bce}{\begin{center}}
\newcommand{\epi}{\end{picture}}
\newcommand{\ece}{\end{center}}
\newcommand{\sumi}{\sum_i}

\newcommand{\lambdar}{\lambda_{\mathrm R}}
\newcommand{\mr}{m_{\mathrm R}}
\newcommand{\xir}{\xi_{\mathrm R}}
\newcommand{\dxir}{\Delta\xi_{\mathrm R}}
\newcommand{\xr}{x_{\mathrm R}}

\newcommand{\qslush}{q\hspace{-0.19cm}/}
\newcommand{\kslush}{k\hspace{-0.23cm}/}

\def\g{{\rm I}\hspace{-0.07cm}\Gamma}

\begin{document}

\begin{flushright}
ECT*-05-01\\[-2mm]
FTUV-2005-01\\[-2mm] 
hep-ph/0501259
\end{flushright}

\title{\Large Displacement Operator Formalism for Renormalization\\[-2mm] 
and Gauge Dependence to All Orders \vspace*{7mm} }


\author{\large Daniele Binosi$^{a,b}$, Joannis Papavassiliou$^c$ and
Apostolos Pilaftsis$^d$ \vspace*{3mm} }

\affiliation{$^a$ECT*, Villa Tambosi, Strada delle Tabarelle 286
I-38050 Villazzano (Trento), Italy \vspace*{1.5mm}\\ 
$^b$I.N.F.N., Gruppo Collegato di Trento, Trento, Italy \vspace*{1.5mm}\\
$^c$Departamento de F\`\i sica Te\`orica, and IFIC
Centro Mixto, 
Universidad de Valencia-CSIC, E-46100, Burjassot, 
Valencia, Spain \vspace*{1.5mm}\\
$^d$School of Physics and Astronomy, University of
Manchester, Manchester M13 9PL, United Kingdom \vspace*{7mm} }

\begin{abstract}
We  present a new method for  determining the renormalization of Green
functions to   all orders in  perturbation  theory, which we  call the
displacement operator formalism, or the $D$-formalism, in short.  This
formalism exploits the fact  that the renormalized Green functions may
be calculated  by  displacing by  an infinite amount  the renormalized
fields and parameters of the theory with respect to the unrenormalized
ones.  With the  help  of this formalism, we   are able to  obtain the
precise form of  the  deformations induced  to the Nielsen  identities
after renormalization,   and thus derive the   exact dependence of the
renormalized   Green  functions  on    the renormalized   gauge-fixing
parameter  to all orders.   As a  particular  non-trivial  example, we
calculate   the gauge-dependence of $\tan\beta$    at two loops in the
framework of an Abelian Higgs  model, using a gauge-fixing scheme that
preserves  the  Higgs-boson low-energy    theorem for  off-shell Green
functions.  Various possible applications   and future directions  are
briefly discussed.
\end{abstract}

\maketitle

\section{Introduction}

Renormalization  is  of central  importance  in  quantum field  theory
\cite{Schwinger:1948iu,Feynman:1948fi,Dyson:1949bp,
Bogoliubov:1957gp,Hepp:1966eg,Zimmermann:1969jj,Symanzik:1970rt,
Epstein:1973gw,'tHooft:1972fi,'tHooft:1971fh,Lee:1974zg,Becchi:1974md},
and its consistent implementation has  been crucial for the advent and
success of  gauge theories  in general, and  of the Standard  Model of
strong  and electroweak  interactions in  particular.  Despite  half a
century  of  practice,  however,  renormalization remains  a  delicate
procedure,  mainly  because   it  interferes  non-trivially  with  the
fundamental  symmetries   encoded  in  the   Lagrangian  defining  the
theory. The  subtleties involved  manifest themselves at  almost every
step,  ranging from  the  necessity to  employ regularization  methods
respecting  all  relevant  symmetries,  the need  for  renormalization
schemes that  do not  spoil the important  constraints imposed  by the
symmetries on  the Green functions of  the theory, all the  way to the
practical, book-keeping challenges  appearing when the renormalization
is carried out in higher order calculations.

In  this paper  we develop  a new  formalism, which  we call  the {\em
Displacement Operator Formalism}, or  the $D$-formalism, in short. The
$D$-formalism  enables one to  systematically organize  and explicitly
compute  the  counterterms   (CTs)  involved  in  the  renormalization
procedure, to  {\it all orders}  in perturbation theory.   The central
observation, leading  to this new  formulation, is that the  effect of
renormalizing any  given Green function  may be expressed in  terms of
ultraviolet (UV) infinite  displacements caused by the renormalization
on  {\it  both}   the  fields  and  the  parameters   of  the  theory.
Specifically, these UV infinite displacements, or shifts, quantify the
difference   between   fields   and   parameters  before   and   after
renormalization. For  example, in the case  of a theory  with a scalar
field $\phi$, and two  parameters, the coupling constant $\lambda$ and
the  squared   mass  $m^2$,  the  corresponding   UV  infinite  shifts
$\delta\phi$,   $\delta\lambda$,  and  $\delta   m^2$  are   given  by
$\delta\phi=(Z_\phi^{\frac12}-1)\phi_{\mathrm{R}}$,
$\delta\lambda=(Z_\lambda-1)\lambdar$,  and  $\delta m^2=(Z_{m^2}-1)\,
m^2_R$,  where the  renormalization constants  are defined,  as usual,
through                       $\phi=Z_\phi^{\frac12}\phi_{\mathrm{R}}$,
$\lambda=Z_\lambda\lambdar$,   and    $m^2=Z_{m^2}\,   m^2_{\rm   R}$.
Evidently, in  this formulation,  dynamical fields and  parameters are
treated on a completely equal footing.

In  order to  systematically  expose  the way  in  which these  shifts
implement  the renormalization at  the level  of Green  functions, one
introduces  the  {\em   displacement  operator}  $D$,  a  differential
operator   given   by   $D   =  \delta\phi   \frac   \partial{\partial
\phi_{\mathrm{R}}}  + \delta\lambda\frac  \partial{\partial\lambdar} +
\delta m^2\frac{\partial}{\partial \mr^2}$.  It turns out (see section
II for details) that the net renormalization effect is captured to all
orders  by  the  exponentiation   of  the  $D$  operator.   Thus,  the
renormalized           ($n$-point)           Green           functions
$\Gamma_{\phi^n}^{\mathrm{R}}$ are  eventually obtained from  the bare
ones,     $\Gamma_{\phi^n}$,    through     the     master    equation
$\phi^n_{\mathrm{R}}\Gamma_{\phi^n}^{\mathrm{R}}    =    \langle   e^D
(\phi^n_{\mathrm{R}}\Gamma_{\phi^n})\rangle$,   where   the   brackets
$\langle\dots\rangle$ means that the corresponding shifts, implicit in
the $D$ operator, are to be  replaced by their expressions in terms of
the renormalization constants given above,  {\it after} the end of the
differentiation procedure.

It   is  important   to  appreciate   at  this   point   the  inherent
non-perturbative nature  of the above  formulation, manifesting itself
through the  exponentiation of the $D$  operator. Perturbative results
(at arbitrary  order) may  be recovered as  a special case  through an
appropriate  order-by-order  expansion of  the  above master  formula,
whose  validity   however  is  not  restricted  to   the  confines  of
perturbation theory.   This is to be contrasted  with other well-known
renormalization methods  (as, for example,  the ``algebraic approach''
\cite{Piguet:1995er,Kraus:1997bi,Grassi:1999tp}), which are formulated
at the perturbative level.

One  of the  main advantages  of this  formulation is  the  ability it
offers  in determining  unambiguously the  CTs to  any given  order in
perturbation theory.   Specifically, the contributions of  the CTs are
automatically obtained through  the straightforward application of the
$D$ operator on the  unrenormalized Green functions, without having to
resort to any additional arguments whatsoever. This last point is best
appreciated in the context  of complicated non-Abelian gauge theories,
or special gauge-fixing schemes, where keeping track of the CTs can be
not  only  logistically  demanding,  but at  times  also  conceptually
subtle. For example, possible  ambiguities related to the gauge-fixing
parameter (GFP) renormalization, endemic in sophisticated quantization
schemes    such    as    the    Background    Field    Method    (BFM)
\cite{Dewitt:ub,'tHooft:vy,Abbott:1980hw,Capper:1982tf,Abbott:zw}  are
automatically resolved (see section III).  As we will see in detail in
the main body of the  paper, the fact that the $D$-formalism correctly
incorporates the action of the CTs is clearly reflected in the absence
of {\it overlapping divergences} from the resulting expressions.

An  important property  of  the $D$-formalism  is  that it  reproduces
exactly the usual diagrammatic representation  of the CTs, if one acts
with  the $D$  operator on  the Feynman  graphs determining  the given
Green  function {\it  before} the  integration over  the  virtual loop
momenta  is carried  out.   If one  instead  acts with  $D$ after  the
momentum  integration  has  been   performed  one  loses  this  direct
diagrammatic interpretation,  but recovers  the same final  answer for
the renormalized Green function.  This last point becomes particularly
relevant in  the case of gauge  theories, where one  of the parameters
that  undergoes   renormalization  is  the  GFP,  to   be  denoted  by
$\xi$. This, in turn, will introduce in the $D$ operator a term of the
form  $\delta\xi\frac\partial{\partial\xi_{\mathrm{R}}}$,  which  must
act on the  corresponding Green function. Clearly, for  this to become
possible the  dependence of the Green  function on $\xi$  must be kept
arbitrary, that is,  one may not choose a  convenient value for $\xi$,
like for  example $\xi=1$. Evidently, if  one were to  first carry out
the momentum  integration and then  differentiate, one would  be faced
with  the book-keeping  complications of  computing with  an arbitrary
$\xi$.   If, instead,  one opts  for the  action of  $D$ on  the Green
function before  the loop integration, the straightforward  use of the
chain  rule for the  strings of  tree-level propagators  (and possibly
vertices), and  a subsequent choice  of, say, $\xi =1$,  is completely
equivalent to the conventional approach of computing at a fixed gauge.
Obviously,  one may  choose either  one  of the  two procedures,  {\it
i.e.}, integrate before  or after the action of  $D$, depending on the
specific problem at hand, a  fact which exemplifies the versatility of
the new formalism.

Perhaps  one  of  the  most  powerful features  of  the  $D$-formalism
consists in  the fact that it  provides a handle on  the structure and
organization of  the CTs to  {\it all orders} in  perturbation theory.
In particular, it  furnishes the {\it exact form}  of the deformations
induced due to the renormalization  procedure to any type of relations
or constraints  which are valid  at the level of  unrenormalized Green
functions.  Such deformations will  arise in general if the symmetries
(global or  local) enforcing the relations in  question involve static
parameters,  which {\it  cannot} admit  kinetic terms  compatible with
those symmetries.  A well-known example  of such a type of relation is
the so-called Nielsen identities (NI) \cite{Nielsen:1975ph}, which, by
virtue  of the {\it  extended} Becchi--Rouet--Stora  ($e$BRS) symmetry
\cite{Kluberg-Stern:1974rs,Piguet:1984js},  control the GFP-dependence
of bare Green function in a completely algebraic way.  The NI are very
useful in  unveilling in their  full generality the patterns  of gauge
cancellations taking  place inside gauge  independent quantities, such
as    $S$-matrix    elements    \cite{Becchi:1974md,Aitchison:1983ns}.
Contrary to  the case of the usual  BRS symmetry \cite{Becchi:1976nq},
however, where only dynamical fields are involved, the $e$BRS symmetry
involves  also  static  parameters  ($\xi$, $\eta$)  which  cannot  be
promoted  to   dynamical  variables  without   violating  this  latter
symmetry.   Consequently, the  NI  are affected  non-trivially by  the
ensuing  renormalization,  which deforms  them  in  a complicated  way
\cite{Gambino:1999ai}.  As  we will see  in detail in section  IV, the
straightforward application of the $D$-formalism yields, for the first
time,  the deformation  of  the NI  in  a closed,  and, in  principle,
calculable  form.   This is  a  clear  improvement  over the  existing
attempts  in  the  literature,  where the  deformations  are  formally
accounted for,  and their  generic structure and  properties inferred,
but,  to  the  best  of  our knowledge,  no  well-defined  operational
procedure for  their systematic computation has been  spelled out thus
far.

The  paper is organized  as follows:  In section  II we  introduce the
$D$-formalism in the context of  a scalar $\phi^4$ theory, and explain
its most characteristic features. The generalization of this formalism
to more complicated field theories  is presented, and a general master
equation valid for  any Green function is derived.   In section III we
apply the  $D$-formalism to  the case of  known examples, in  order to
familiarize  the reader  with its  use.  In  particular, we  study the
two-loop  renormalization for  the  cases of  $\phi^4$,  QED, and  QCD
formulated  in the  BFM, and  demonstrate  how all  necessary CTs  are
unambiguously generated through the straightforward application of the
$D$  operator.   Section  IV  contains the  first  highly  non-trivial
application  of  the  $D$-formalism,  namely  the  derivation  of  the
deformation induced to the NI by the renormalization procedure. First,
we  review  the derivation  of  the  NI,  and introduce  a  consistent
graphical  representation.  Then,  by means  of the  $D$-formalism, we
obtain  for the first  time the  NI deformation  in a  precise, closed
form, presented in Equation~(\ref{dNI}),  which is a central result of
this paper.

In    section    V    we    revisit   the    Abelian    Higgs    model
\cite{Kastening:1993zn},  and study  its quantization  in  the special
class  of gauges  which preserve  the Higgs-boson  Low  Energy Theorem
(HLET)~\cite{Ellis:1975ap,Shifman:1978zn,Shifman:1979eb,Vainshtein:1980ea,
Dawson:1989yh} beyond the tree-level.   The reason for turning to this
model is three-fold. First, it  has a rich structure, most notably two
different GFPs, thus  providing an ideal testing ground  for the newly
introduced formalism.   Second, in theories  with spontaneous symmetry
breaking (SSB), the  HLET is of central importance,  and together with
the Slavnov-Taylor  identities \cite{Slavnov:1972fg} and  NI, severely
constrains  the  form  of  the  Green functions  of  the  theory.   By
quantizing    the    model    using   the    HLET-preserving    gauges
\cite{Pilaftsis:1997fe} we allow  for the consistent implementation of
a  very  special  renormalization   condition  for  the  Higgs  vacuum
expectation value  (VEV) $v$. Specifically,  if the gauge  fixing does
not preserve  the HLET, as is,  for example, the case  of the ordinary
$R_{\xi}$ gauges \cite{Fujikawa:fe}, the renormalization of $v$ is not
multiplicative   in   terms   of   the   Higgs-boson   wave   function
renormalization  $Z_H$,  but  necessitates an  additional  (divergent)
shift,  $\Delta  v$, {\it  i.e.},~$v  = Z_H^\frac{1}{2}(v_{\mathrm  R}
+\Delta  v)$  \cite{Bohm:1986rj,Chankowski:1991md}.   Instead, in  the
HLET-preserving gauges, only the Higgs wavefunction $Z_H$ is needed to
renormalize $v$, while  the shift $\Delta v$ is  {\it finite} ($\Delta
v$ vanishes  in the  $\overline{\rm MS}$ scheme  \cite{MSbar}).  Being
able to  renormalize $v$ only  by means of  $Z_H$ is essential  in the
study of the gauge-(in)dependence of $\tan\beta$ in the context of the
HLET-preserving gauges, appearing in the  next section, as well as the
definition of  effective charges  for the Higgs  sector of  the theory
\cite{Papavassiliou:1997fn}.   Third,  by  including fermions  in  the
model, we set up the stage for the calculations that will be presented
in the next section.

In  section VI we  address the  issue of  the gauge-dependence  of the
$\tan\beta$,  one  of  the  most  important  parameters  appearing  in
multi-Higgs  models  \cite{Gunion:1989we,Haber:1978jt}.   We  show  by
means of explicit two-loop calculations that the condition of having a
vanishing  $\Delta  v$  does  not guarantee  the  GFP-independence  of
$\tan\beta$,  because the  difference of  the  corresponding anomalous
dimensions displays an explicit dependence  on the GFP.  This is to be
contrasted   to   what  happens   in   the   context   of  the   usual
(non-HLET-preserving)   $R_{\xi}$  gauges,  where   the  corresponding
difference  is GFP-independent,  but the  GFP-dependence  enters again
into    $\tan\beta$    through    the   non-vanishing    $\Delta    v$
\cite{Yamada:2001ck,Freitas:2002um}.  In the calculations presented in
this  section  we make  extensive  use  of  the $D$-formalism,  which,
together with the HLET, simplifies substantially the algebra involved.
Our conclusions and outlook are discussed in section VII.  Finally, in
an Appendix  we list  the Feynman rules  for the Abelian  Higgs model,
together  with a  collection  of  results needed  for  several of  the
calculations appearing in this article.

\section{The displacement operator formalism}

In  this  section we  present  the  derivation  of the  aforementioned
$D$-formalism. For this pourpose it is sufficient to consider a simple
scalar  $\lambda\phi^4$ theory;  the generalization  to  more involved
field theories  (such as gauge theories) is  straightforward, and will
be done at the end of this section.

Let then  $\Gamma_{\phi^n}$ be a bare,  one-particle irreducible (1PI)
$n$-point  Green  function.  After  carrying  out the  renormalization
programme, in $d=4-\epsilon$ dimensions, one will have that
\begin{equation}
  \label{Gren}
\phi^n\Gamma_{\phi^n}(\lambda,m^2;\mu,\epsilon)\ =\ \phi^n_{\mathrm{R}}
\Gamma_{\phi^n}^{\mathrm{R}}(\lambdar,\mr^2;\mu)\; , 
\end{equation}
where   $\lambda,  m^2,     \phi$  (respectively  $\lambdar,    \mr^2,
\phi_{\mathrm{R}}$) are the bare (respectively renormalized) parameters
and dynamical field   of the theory   at hand~\cite{footnote1}, 
\begin{equation}
\phi=Z_\phi^{\frac12}\phi_{\mathrm{R}}, \qquad
\lambda=Z_\lambda\lambdar, \qquad m^2 = Z_{m^2}\mr^2, 
\label{CTs}
\end{equation} 
and    $\Gamma_{\phi^n}^{\mathrm{R}}(\lambdar;\mu)$   represents   the
renormalized $n$-legs Green function.  Note that our definition of the
bare  parameters $\phi$  and $\lambda$  does not  include  their naive
dimensional  scalings, namely $\mu^{-\epsilon/2}$  and $\mu^\epsilon$,
respectively.   Then,  we may  rewrite  the  left-hand  side (LHS)  of
(\ref{Gren}) as
\begin{equation} 
\phi^n\Gamma_{\phi^n}(\lambda,m^2;\mu,\epsilon)\ =\
(\phi_{\mathrm{R}}+\delta\phi)^n
\Gamma_{\phi^n}(\lambdar+\delta\lambda,\mr^2 + \delta m^2;\mu,\epsilon)\;, 
\end{equation} 
where the parameter shifts 
\begin{equation}
\delta\phi=(Z_\phi^{\frac12}-1)\phi_{\mathrm{R}}, \qquad
\delta\lambda=(Z_\lambda-1)\lambdar, \qquad \delta
m^2=(Z_{m^2}-1)\mr^2, 
\label{shmore}
\end{equation} 
quantify the difference of the renormalized quantities with respect to
the  corresponding  unrenormalized  ones  in a  given  renormalization
scheme R. Notice that our formulation of renormalization treats fields
and couplings on equal footing, {\it i.e.}, as independent fundamental
parameters of the theory.

Our next step  is to use a Taylor expansion  to trade the combinations
$\phi_{\mathrm{R}}+\delta\phi$,      $\lambdar+\delta\lambda$,     and
$\mr^2+\delta m^2$,  for the renormalized  parameters $\phi_{\mathrm{R}}$,
$\lambdar$, and  $\mr$.  To this end,  it is natural  to introduce the
differential displacement operator
\begin{equation}
D\ =\ \delta\phi\frac\partial{\partial\phi_{\mathrm{R}}}+
\delta\lambda\frac\partial{\partial\lambdar}+\delta
m^2\frac{\partial}{\partial \mr^2}\; ,
\end{equation}
in which the shifts are treated as independent parameters, and only at
the very end of all the manipulations they will be replaced by their
actual values in terms of the renormalization constants
$Z_\phi(\lambdar,\mr^2;\epsilon)$,
$Z_\lambda(\lambdar,\mr^2;\epsilon)$, 
and $Z_{m^2}(\lambdar,\mr^2;\epsilon)$.  It is then not difficult to derive
the following (all-order) master equation
\begin{equation}
  \label{master}
(\phi_{\mathrm{R}}+\delta\phi)^n\,\Gamma_{\phi^n}(\lambdar +
\delta\lambda,\mr^2+\delta m^2;\mu,\epsilon) \ =\
\Big<\,e^D\,\phi^n_{\mathrm{R}}\,
\Gamma_{\phi^n}(\lambdar,\mr^2;\mu,\epsilon)\, \Big>\;,
\end{equation}
or equivalently            
\begin{equation}
  \label{master2}
\phi^n_{\mathrm{R}}\,
\Gamma_{\phi^n}^{\mathrm{R}}(\lambdar,\mr^2;\mu)\ =\ 
\Big<\,e^D\,\phi^n_{\mathrm{R}}\,
\Gamma_{\phi^n}(\lambdar,\mr^2;\mu,\epsilon)\, \Big>\;, 
\end{equation}
where   $\langle  \dots   \rangle$  means   that  the   CT  parameters
$\delta\phi$,  $\delta \lambda$,  and $\delta  m^2$ are  to be  set to
their actual  values according to  (\ref{shmore}) after the  action of
the $D$  operator. The expression~(\ref{master2}),  being an all-order
result, can now be expanded to any given order in perturbation theory.
This means that not only $\Gamma_{\phi^n}$ should be expanded starting
from  tree  level, but,  accordingly,  also  the shifts  $\delta\phi$,
$\delta\lambda$  and  $\delta  m^2$,  together with  the  displacement
operator $D$ (in the latter  case the expansion starts at the one-loop
level, since the tree level shifts are, of course, zero).

1Now, since the shifts are treated as independent parameters, the
displacement operator at different perturbative orders commute, and
one can use the ordinary Taylor expansion for the exponentiation of
$D$; up to three loops, one then gets
\begin{equation}
e^D\ =\ 1\: +\: D^{(1)}\: +\: \Big( D^{(2)} +
\frac12D^{(1)\,2}\Big)\: +\: \Big(
D^{(3)}+D^{(2)}D^{(1)}+\frac16D^{(1)\,3}\Big)\: +\: \dots,
\end{equation}
with
\begin{equation}
D^{(n)}\ =\ \delta\phi^{(n)}\frac{\partial}{\partial\phi_{\mathrm R}} +
\delta\lambda^{(n)}\frac{\partial}{\partial\lambdar} + 
\delta m^{2\,(n)}\frac{\partial}{\partial\mr^2}\ .
\end{equation}
The  parameter  shifts   $\delta\phi^{(n)}$, $\delta\lambda^{(n)}$, and
$\delta m^{2\,(n)}$ are loop-wise defined as follows:
\begin{equation}
\delta \phi^{(n)}\ =\ Z^{\frac12\, (n)}_\phi\, \phi_{\rm R}\,,\qquad
\delta\lambda^{(n)}\ =\ Z^{(n)}_\lambda\, \lambda_{\rm R}\,,\qquad
\delta m^{2\,(n)}\ =\ Z^{(n)}_{m^2}\, m^2_{\rm R}\; .
\end{equation}
By acting  with the  operator $e^D$ on  the 1PI  correlation functions
$\phi^n_{\mathrm{R}}\Gamma_{\phi^n}(\lambdar,\mr^2;\mu,\epsilon)$,   we
can easily determine the  expressions for the renormalized correlation
functions                                          $\phi^n_{\mathrm{R}}
\Gamma_{\phi^n}^{\mathrm{R}}(\lambdar,\mr^2;\mu)$  at  the one-,  two-
and three-loop level, which read
\begin{eqnarray}
  \label{exp} \phi^n_{\mathrm{R}}
\Gamma_{\phi^n}^{\mathrm{R}(1)}(\lambdar,\mr^2;\mu)&=&
\Big<\,D^{(1)}\phi^n_{\mathrm{R}}\Gamma_{\phi^n}^{(0)}(\lambdar,\mr^2;\mu)\,
\Big>\: +\:
\phi^n_{\mathrm{R}}\Gamma_{\phi^n}^{(1)}(\lambdar,\mr^2;\mu,\epsilon)\;,
\nonumber \\[3mm] \phi^n_{\mathrm{R}}
\Gamma_{\phi^n}^{\mathrm{R}(2)}(\lambdar,\mr^2;\mu)&=& \Big<\,\Big(
D^{(2)}+\frac12D^{(1)\,2} \Big)\,\phi^n_{\mathrm{R}}
\Gamma_{\phi^n}^{(0)}(\lambdar,\mr^2;\mu)\: +\: D^{(1)}
\phi^n_{\mathrm{R}}\Gamma_{\phi^n}^{(1)}(\lambdar,\mr^2;\mu,\epsilon)\,\Big>
\nonumber\\
&&+\,\phi^n_{\mathrm{R}}\Gamma_{\phi^n}^{(2)}(\lambdar,\mr^2;\mu,\epsilon)\;,
\nonumber\\[3mm] \phi^n_{\mathrm{R}}
\Gamma_{\phi^n}^{\mathrm{R}(3)}(\lambdar,\mr^2;\mu)&=& \Big<\,
\Big(\,D^{(3)}+D^{(2)}D^{(1)}+\frac16D^{(1)\,3}\,\Big)
\phi^n_{\mathrm{R}}\Gamma_{\phi^n}^{(0)} (\lambdar,\mr^2;\mu)
\nonumber \\ &&+\,
\Big(D^{(2)}+\frac12D^{(1)\,2}\Big)\,\phi^n_{\mathrm{R}}
\Gamma_{\phi^n}^{(1)}(\lambdar,\mr^2;\mu,\epsilon)\: +\: 
D^{(1)} \phi^n_{\mathrm{R}}\Gamma_{\phi^n}^{(2)}
(\lambdar,\mr^2;\mu,\epsilon)\,\Big>\nonumber\\
&&+\,\phi^n_{\mathrm{R}}\Gamma_{\phi^n}^{(3)}(\lambdar,\mr^2;\mu,\epsilon)\; .
\end{eqnarray}

It  is easy  to generalize  the  above formalism  to more  complicated
theories,  such as  gauge  theories  (with or  without  SSB).  Let  us
consider a field theory, where  the dynamical degrees of freedom, {\it
e.g.}, scalar, spinor  or vector fields, are all  denoted by $\phi_i$,
while the static parameters of  the theory, such as couplings, masses,
VEVs   of    fields,   and   GFPs,    are   collectively   represented
by~$x$. Correspondingly,  let us assume that the  dynamical and static
parameters renormalize multiplicatively as follows:
\begin{equation}
\phi_i = Z^{\frac12}_{\phi_i}\phi_{i\mathrm R}, \qquad
x = Z_x x_{\mathrm R}.
\label{genth}
\end{equation}
For such  a  theory, the   displacement  operator  $D$ will  be  given
by~\cite{footnote2}
\begin{equation}
D = \sum_i\delta\phi_i\frac\partial{\partial\phi_{i\mathrm{R}}}+
\sum_x\delta x\frac\partial{\partial x_{\mathrm{R}}}\ ,
\end{equation}
where the CT shifts are defined according to the loop-wise expansion
\begin{equation}
  \label{CTshift}
\delta\phi_i\ =\ (Z^{\frac12}_{\phi_i}-1)\phi_{i\mathrm{R}}\ =\
\sum_{n=1}^\infty Z^{\frac12\,(n)}_{\phi_i}\, \phi_{iR}\,, \qquad
\delta x\ =\ (Z_x-1)x_{\mathrm{R}}\ =\ \sum_{n=1}^\infty\, Z^{(n)}_x\, x_R\;.
\end{equation}
Then, given any  1PI Green function $\Gamma_{\prod_i\phi_i^{n_i}}$, it
will renormalize according to the formula
\begin{equation}
  \label{Dformalism}
\prod_i\,\phi_{i\mathrm R}^{n_i}\,\Gamma_{\prod_i\phi_i^{n_i}}^{\mathrm
  R}(x_{\mathrm R};\mu)\ =\
\Big<\, e^D\, \prod_i\,\phi_{i{\rm R}}^{n_i}\,
\Gamma_{\prod_i\phi_i^{n_i}}(x_{\mathrm R};\mu,\epsilon)\,\Big>\; .
\end{equation}
Equation~(\ref{Dformalism}) is the master formula of the $D$-formalism
and one of  the major results of this  paper. An immediate consequence
of the $D$-formalism is the renormalization-scheme independence of the
LHS  of~(\ref{Dformalism}).  Specifically,  if R  and R$'$  denote two
arbitrary   renormalization    schemes,   the   renormalization-scheme
independence of the bare parameters, {\it i.e.},~$\phi_i = \phi_{i{\rm
R}} +  \delta\phi^{{\rm (R)}}_i  = \phi_{i{\rm R'}}  + \delta\phi^{\rm
(R')}_i$ and $x = x_{\rm R} + \delta x^{\rm (R)} = x_{\rm R'} + \delta
x^{({\rm R'})}$, implies that
\begin{equation}
\prod_i\phi_{i\mathrm R}^{n_i}\Gamma_{\prod_i\phi_i^{n_i}}^{\mathrm
  R}(x_{\mathrm R};\mu)=
\prod_i\phi_{i\mathrm R'}^{n_i}\Gamma_{\prod_i\phi_i^{n_i}}^{\mathrm
  R'}(x_{\mathrm R'};\mu).
\end{equation}
Thus, we may  employ the $D$-formalism to go from the  scheme R to the
scheme R$'$,  and so generally establish the  precise relation between
the  respective  1PI  Green   functions  evaluated  in  two  different
renormalization schemes.

\section{Examples}

To  get  acquainted with  this  new  renormalization  method, in  this
section  we are  going to  reproduce  known results,  and clarify  the
connection   between   the   $D$   operator   and   the   conventional
renormalization procedure.

\subsection{The case of $\lambda\phi^4$ }
\begin{figure}[t]
\bce
\includegraphics[width=15cm]{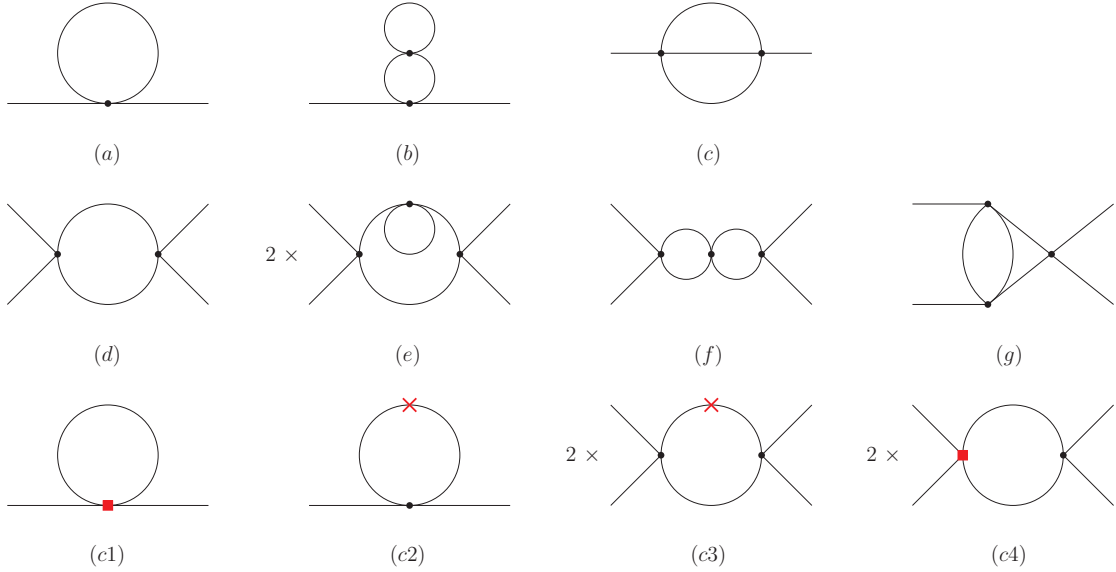}
\caption{\label{lphi4} One and two-loop propagator and vertex
corrections in the $\lambda\phi^4$ theory. Dots represents $\phi^4$
vertices, while crosses and squares represent the one-loop propagator
and vertex counterterms respectively. Crossing diagrams are not
shown.}  \ece
\end{figure}

Let  us start by  first considering  the usual  $\lambda\phi^4$ theory
defined by
\begin{equation}
{\cal L}\ =\ \frac12(\partial^\mu\phi)(\partial_\mu\phi)\: -\: 
\frac12 m^2\phi^2 \: +\: \frac\lambda{4!}\phi^4\; .
\end{equation}
At the  one-loop  level, the first  equation of (\ref{exp})  gives the
usual renormalization prescription in the MS scheme,
\begin{eqnarray}
Z^{(1)}_\phi&=& 2Z^{\frac12\,(1)}_\phi\nonumber\\
&=& i\frac\partial{\partial p^2} 
\overline{\Gamma_{\phi^2}^{(1)}(\lambdar,\mr^2;\mu,\epsilon)}_{|_{p^2
= 0}}\; ,\nonumber\\ 
Z^{(1)}_\lambda&=&\frac i\lambdar
\overline{\Gamma_{\phi^4}^{(1)}(\lambdar,\mr^2;\mu,\epsilon)}\; ,\nonumber\\
Z^{(1)}_\phi+Z^{(1)}_{m^2}&=&-\frac i{\mr^2}
\overline{\Gamma_{\phi^2}^{(1)}(\lambdar,\mr^2;\mu,\epsilon)}_{|_{p^2=0}}\; ,
\end{eqnarray}
where  the  ``bars'' indicate  that  only  the  infinite part  of  the
corresponding   bare   Green  function   should   be  considered.    A
straightforward    calculation   of    the   one-loop    diagrams   of
Fig.~\ref{lphi4} (with $d=4-\epsilon$) gives then
\begin{eqnarray}
Z^{(1)}_\phi=0\,,\qquad
Z^{(1)}_{m^2}=-\frac1{(4\pi)^2\epsilon}\lambda\,, \qquad
Z^{(1)}_\lambda=-\frac1{(4\pi)^2\epsilon}3\lambda\;.
\end{eqnarray}
More interesting is  the two-loop case, for which  the second equation
in (\ref{exp}) gives
\begin{eqnarray}
Z^{(2)}_\phi&=& 2Z^{\frac12\,(2)}_\phi\: +\: 
\left(Z^{\frac12\,(1)}_\phi\right)^2\ \nonumber \\
&=&i\frac\partial{\partial p^2}\left[\
\overline{D^{(1)}\Gamma_{\phi^2}^{(1)}(\lambdar,\mr^2;\mu,\epsilon)}+
\overline{\Gamma_{\phi^2}^{(2)}(\lambdar,\mr^2;\mu,\epsilon)}\
\right]_{|_{p^2=0}},\nonumber\\ 
Z^{(2)}_\lambda+Z^{(2)}_\phi+Z^{(1)2}_\phi+2Z^{(1)}_\phi
Z^{(1)}_\lambda&=&\frac i{\lambdar}\left[\
\overline{D^{(1)}\Gamma_{\phi^4}^{(1)}(\lambdar,\mr^2;\mu,\epsilon)}+
\overline{\Gamma_{\phi^4}^{(2)}(\lambdar,\mr^2;\mu,\epsilon)}\
\right],\nonumber\\ 
Z^{(2)}_{m^2}+Z^{(2)}_\phi+Z^{(1)}_\phi Z^{(1)}_{m^2}&=&-\frac
i{\mr^2}\left[\
\overline{D^{(1)}\Gamma_{\phi^2}^{(1)}(\lambdar,\mr^2;\mu,\epsilon)}+ 
\overline{\Gamma_{\phi^2}^{(2)}(\lambdar,\mr^2;\mu,\epsilon)}\
\right]_{|_{p^2=0}},\nonumber\\ 
\end{eqnarray}
where  $D^{(1)}$ is  now  the displacement  operator  {\it after}  the
one-loop parameters shift has been substituted, {\it i.e.},
\begin{equation}
D^{(1)}\ =\ 
Z^{(1)}_\lambda\lambdar\frac\partial{\partial\lambdar}\: +\: Z^{(1)}_{m^2}
\mr^2\frac\partial{\partial\mr^2}\: +\: \frac n2Z^{(1)}_\phi\; ,
\end{equation}
with  $n$ is the  number of  external $\phi$  legs. Notice  that these
equations  tell us  a rather  non-trivial fact,  {\it i.e.},  that the
combination in square brackets on the right-hand side (RHS) is free of
overlapping  divergences.   This  is  quite  striking,  since  in  the
calculation  of the  two loop  Green  function $\Gamma_{\phi^n}^{(2)}$
only 1PI bare diagrams must be considered (see Fig.~\ref{lphi4} again)
and no two-loop  CT diagram has to be  taken into account.  Evidently,
the  $D^{(1)}$ operator, when  applied to  the lower  order (one-loop)
$\Gamma^{(1)}_{\phi^n}$,  will  generate   the  necessary  CT  at  the
two-loop order.

Since  the  $D$ operator  and  the  integration  over virtual  momenta
commute, one  has two possibilities:  first carry out  the integration
and then  apply $D$, or,  {\it vice versa},  first apply $D$  and then
integrate.   These   two   approaches   have   both   advantages   and
disantvantages.  In  the first  case, the main  advantage is  that all
explicit references  to CTs is removed,  and one needs  to compute 1PI
irreducible  diagrams  only (no  CT  diagrams).   The disadvantage  is
related   to  the  fact   that,  since   the  $D$   operator  involves
differentiation  with respect to  {\it all}  parameters of  the theory
that  undergo  renormalization,  we  need  to  maintain  the  explicit
dependence  of   the  various  Green  functions  on   them.   This  is
particularly relevant in  the case of gauge theories,  where one would
be faced with  the complication of computing diagrams  keeping the GFP
$\xi$ arbitrary  \cite{footnote3}.  In  the second case,  instead, one
will  algebraically  generate the  CTs  and  recover the  conventional
formulation. Of course, after applying  the $D$ operator, one can work
at  a fixed  $\xi$ (say,  $\xi=1$).  To  see how  the CT  diagrams get
generated by  the displacement operator, observe that  the one-loop CT
to be added to the Lagrangian are given by
\begin{equation}
\delta c_1=\frac i{(4\pi)^2\epsilon}\lambda m^2, \qquad 
\delta c_2=-\frac i{(4\pi)^2\epsilon}3\lambda^2,
\end{equation}
which are related to the renormalization of the operators $\phi^2$ and
$\phi^4$,  respectively.  Then, by  applying the  $D$ operator  on the
bare one-loop 1PI diagrams $(a)$  and $(d)$ of Fig.~\ref{lphi4}, it is
easy to see that
\begin{eqnarray}
D^{(1)}\Gamma_{\phi^2}^{(1)}&=&-i\delta c_1\frac\lambda2\int\!
\frac{d^dk}{(2\pi)^d} \frac1{(k^2-m^2)^2}+i\delta c_2\frac12\int\!
\frac{d^dk}{(2\pi)^d}\frac1{k^2-m^2}\nonumber \\ &=&(c1)+(c2),\nonumber \\ 
D^{(1)}\Gamma_{\phi^4}^{(1)}&=&3i\delta
c_1\lambda^2\int\! \frac{d^dk}{(2\pi)^d} \frac1{(k^2-m^2)^3}-3i\delta
c_2\lambda\int\! \frac{d^dk}{(2\pi)^d}\frac1{(k^2-m^2)^2}\nonumber \\
&=&(c3)+(c4).
\end{eqnarray}
Therefore, irrespectively of the order, one will get the results
\begin{eqnarray}
Z^{(2)}_\phi=-\frac1{(4\pi)^4}\frac{\lambda^2}{12\epsilon},\quad
Z^{(2)}_{m^2} = 
-\frac1{(4\pi)^4}\lambda^2\left(\frac2{\epsilon^2} 
-\frac5{12\epsilon}\right),\quad
Z^{(2)}_\lambda = -\frac1{(4\pi)^4}\lambda^2 
\left(\frac9{\epsilon^2}-\frac{35}{12\epsilon}\right).\nonumber\\ 
\end{eqnarray}

\subsection{Massless QED}

\begin{figure}[t]
\bce
\includegraphics[width=10cm]{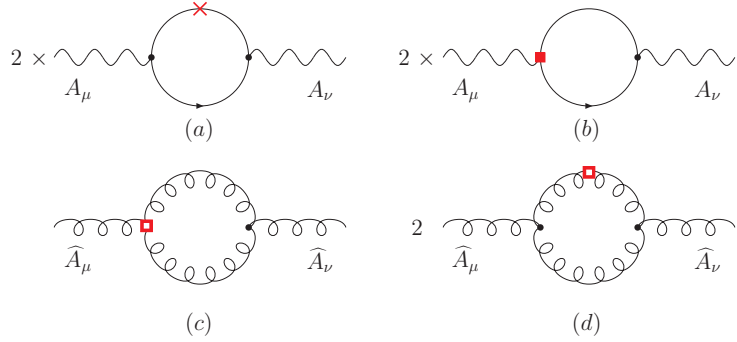}
\caption{\label{QEDandQCD} One-loop CTs for massless QED
[($a$) and ($b$)], and pure Yang-Mills theory in the background field
type of gauges [($c$) and ($d$)]. In the latter case we only draw
gauge-fixing term insertions resulting from the
renormalization of the GFP.} 
\ece
\end{figure}

We turn to the case of  QED, where for simplicity the electron mass is
considered  to be  zero to  all orders.   In this  case  the one-loop 
$D$ operator  of the photon vacuum polarization, assumes   the    form 
\begin{equation}
D^{(1)}\ =\ Z^{(1)}_e e_R\frac  \partial{\partial e_R}+Z^{(1)}_{\xi} \xi_R
\frac\partial{\partial\xi_R}+  Z^{(1)}_A\; ,  
\end{equation}
where  $e$  is the  QED  coupling,  and  $Z^{(1)}_e$, $Z^{(1)}_A$  and
$Z^{(1)}_{\xi}$   are,   respectively,    the   one   loop   coupling,
wave-function,  and GFP  renormalization constants.
It  is  then straightforward  to  establish  that  the action  of  the
operator  $D^{(1)}$   on  the  one-loop   photon  vacuum  polarization
$\Gamma_{A_{\mu}A_{\nu}}^{(1)}$ vanishes.   To begin with,  the vacuum
polarization is independent of the GFP to all orders. In addition, the
Abelian gauge  symmetry of  the theory gives  rise to  the fundamental
Ward  identity  
\begin{equation}  q^{\mu} \Gamma^0_{\mu}(p,p+q)=
S_0^{-1}(p+q)-S_0^{-1}(p)\; ,  
\end{equation} 
where   $\Gamma^0_{\mu}$  and   $S_0  (k)$   are   the  unrenormalized
(all-order)   photon-electron    vertex   (with an $e$ factored out)
and   electron   propagator,
respectively.    The   requirement   that  the   renormalized   vertex
$\Gamma_{\mu} = Z_{1}\Gamma^0_{\mu}$  and the renormalized self-energy
$S =  Z_{f}^{-1} S_0$ satisfy  the same identity imposes  the equality
$Z_{1}=Z_{f}$,     from     which  follows immediately     that
$Z_{e}=Z_{A}^{-1/2}$,  and therefore, after  expanding perturbatively,
$2 Z^{(1)}_e + Z^{(1)}_A = 0$. Then, using that, to the given order,
\begin{equation} 
e_R \frac \partial {\partial e_R} \Gamma_{A_{\mu}A_{\nu}}^{(1)}\ =\
2 \Gamma_{A_{\mu}A_{\nu}}^{(1)}\; , 
\end{equation}
we find immediately that $D^{(1)}\Gamma_{A_{\mu}A_{\nu}}^{(1)}  = 0$.
Evidently,   the  action   of  $D^{(1)}$   produces  no   two-loop  CT
contributions. This is  consistent with the fact that  the standard CT
graphs (first  row in Fig.~\ref{QEDandQCD})  add up to zero,  again by
virtue    of    the   aforementioned    Ward    identity   (see    for
example~\cite{Itzykson:1980rh}).

At   higher  orders,   the  renormalization   of  the   photon  vacuum
polarization   requires   CTs  that   originate   from  two-loop   and
higher-order  self-energy graphs.   Likewise, it  is not  difficult to
show that, on account of the relation $Z_e = Z^{-1/2}_A$, the CTs from
one-loop graphs vanish identically to all orders.  In QED with massive
quarks  and leptons,  mass  CTs from  all  lower-order vacuum  graphs,
including one-loop  graphs, contribute  to the renormalization  of the
photon vacuum polarization.  In this case, the $D$-formalism is a very
practical method to reliably calculate such effects.

\subsection{Gluon self-energy in the Background Field Method}

We next turn to the case of quark-less QCD, formulated in the BFM, and
apply the  $D$-formalism to  study the structure  of the  two-loop CTs
appearing in  the calculation of the background  gluon self-energy, to
be  denoted by  $\Gamma_{\widehat {A}_{\mu}  \widehat{A}_{\nu}}$.  The
BFM is a special  gauge-fixing procedure, which preserves the symmetry
of the action under ordinary gauge transformations with respect to the
background  (classical) gauge  field $\widehat{A}^a_{\mu}$,  while the
quantum  gauge fields  $A^a_{\mu}$  appearing in  the loops  transform
homogeneously under the gauge  group \cite{Weinberg:kr}.  As a result,
the     $n$-point    functions     $\langle    0     |     T    \left[
\widehat{A}^{a_1}_{\mu_1}(x_1)      \widehat{A}^{a_2}_{\mu_2}(x_2)\dots
\widehat{A}^{a_n}_{\mu_n}(x_n)  \right]  |0  \rangle$  satisfy  naive,
QED-like  Ward identities, instead  of the  Slavnov--Taylor identities
valid  in the  usual  covariant quantization.   Notice  also that  the
$n$-point  functions  depend  {\it  explicitly}  on  the  quantum  GFP
$\xi_Q$.

In  this case, the one-loop   $D$ operator for the (background)  gluon
vacuum polarization is given by
\begin{equation}
D^{(1)}\ =\ Z^{(1)}_g g_R\frac\partial{\partial g_R}\: +\:
Z^{(1)}_{\xi_Q} \xi_Q \frac\partial{\partial\xi_Q}\: +\:
Z^{(1)}_{\widehat {A}}\; ,
\end{equation}
Here and in  the following, $\xi_Q$ stands for  the {\em renormalized}
GFP associated with the quantum gauge field $A^a_\mu (x)$.  Because of
the    aforementioned   background    gauge   symmetry,    $Z_g$   and
$Z_{\widehat{A}}$ satisfy to all  orders the QED-like relation $Z_{g}\
=\  Z_{\widehat {A}}^{-1/2}$, from  which follows  that the  first and
third term of $D^{(1)}$ cancel  against each other, as happens for the
photon vacuum polarization.  Therefore, $D^{(1)}$ reduces to
\begin{equation}
D^{(1)}\ =\ Z^{(1)}_{\xi_Q} \xi_Q \frac\partial{\partial\xi_Q}.
\label{dxi}
\end{equation}
On the other hand notice that, 
contrary  to the case of the  photon vacuum polarization,
the   background   gluon   self-energy   $\Gamma_{\widehat   {A}_{\mu}
\widehat{A}_{\nu}}$  depends explicitly  on  $\xi_Q$. This  dependence
stems  not  only  from   the  tree-level  (quantum)  gluon  propagator
appearing in the loop,
\begin{equation}
\Delta_{\mu\nu}^{[0]}(k)\ =\ 
-\frac{ i}{k^2}
\left[\ g_{\mu\nu} - (1-\xi_Q) \frac{ k_\mu
k_\nu}{ k^2}\right],
\label{GluProp}
\end{equation}
as happens in  the case of the covariant gauges,  but in addition from
the    tree-level     vertices    between    a     background    gluon
$\widehat{A}^a_{\mu}(p)$  and two  quantum  gluons $A^b_{\nu}(q)$  and
$A^c_{\lambda}(k)$,
\begin{equation}
\Gamma^{abc}_{\mu\nu\rho}(p,q,k)\ =\ g f^{abc} 
\left[ g_{\mu\rho}\left(p- k - \xi_Q^{-1} q\right)_{\nu} + 
g_{\nu\rho}\left( k-q \right)_{\mu}
+ g_{\mu\nu}\left(q- p + \xi_Q^{-1} k \right)_{\rho} \right]\;,
\end{equation}
which, unlike the  usual case, also depend on  $\xi_Q$.  Finally, as a
consequence of the non-renormalization of the longitudinal part of the
gluon     self-energy    to     all    orders,     we     know    that
$Z_{\xi_Q}=Z_{A}$~\cite{Capper:1982tf},    where   $Z_{A}$    is   the
wave-function renormalization of the quantum gluons; it coincides with
the one computed  in the covariant gauges. Thus,  at one-loop (recall that
in our conventions $d=4-\epsilon $)
\begin{equation}
 Z_{\xi_Q}^{(1)}\ =\ \frac{g^2 C_A}{(4\pi)^2}\left(
 \frac{13}{3}-\xi_Q\right)\frac{1}{\epsilon}\;.
\label{xiren}
\end{equation}

Therefore  the  action of  $D^{(1)}$  will  be  non-trivial, and  will
generate     automatically      all     necessary     two-loop     CTs
(Fig.~\ref{QEDandQCD} second  row).  Notice  that this is  in complete
accordance with  the fact  that the only  CTs appearing in  the classic
two-loop calculation of  \cite{Abbott:1980hw} are precisely related to
the gauge-fixing.  The advantage  of the $D$-formalism in this respect
is that the presence  of such CTs does not have to  be deduced, but is
given directly from the very form of the $D$ operator.

In order to  verify that the action of  $D^{(1)}$ on $\Gamma_{\widehat
{A}_{\mu} \widehat{A}_{\nu}}^{(1)}$  generates indeed the  two-loop CT
contributions,  one may  follow two  equivalent ways.  First,  one may
differentiate   directly   on    the   Feynman   diagram   determining
$\Gamma_{\widehat  {A}_{\mu}  \widehat{A}_{\nu}}^{(1)}$,  taking  into
account the dependence of the  propagators and vertices on $\xi_Q$, as
shown above,  and setting  eventually $\xi_Q=1$.  Thus,  one generates
immediately  the  Feynman  graphs  appearing  in  the  second  row  of
Fig.~\ref{QEDandQCD}.

The second, more algebraic way,  is to actually act with the $D^{(1)}$
of   (\ref{dxi})   on   the    expression   that   one   obtains   for
$\Gamma_{\widehat  {A}_{\mu} \widehat{A}_{\nu}}^{(1)}$ after  the loop
integration, and verify that it is actually equal to the sum of the CT
diagrams, whose individual expressions,  calculated at $\xi_Q =1$, are
listed in the Table 1  of \cite{Abbott:1980hw} [where the two diagrams
($c$) and ($d$) are called  ($l$) and ($m$) respectively].  The latter
sum reads
\begin{equation}
(c)+(d)\ =\ \frac{g^4 C_A^2}{(4\pi)^4}
\delta^{ab} \Bigg(q^2 g_{\mu\nu} - q_{\mu}q_{\nu}\Bigg) \frac{20}{3}
\frac{1}{\epsilon}\ .   
\label{sumctBFM}
\end{equation}
In doing so, note that,  as is well known, the $\xi_Q$-dependent part,
$\tilde\Gamma_{\widehat  {A}_{\mu} \widehat{A}_{\nu}}^{(1)}$,  is {\it
finite}, and is given by \cite{Papavassiliou:1994yi}
\begin{equation}
\tilde\Gamma_{\widehat   {A}_{\mu}^a  \widehat{A}_{\nu}^b}^{(1)}(q)\  =\
\frac{g^2   C_A}{(4\pi)^2}    \delta^{ab}   \Bigg(q^2   g_{\mu\nu}   -
q_{\mu}q_{\nu}\Bigg) \frac{(\xi_Q -1)(7+\xi_Q)}{4}\ .
\label{finBFM}
\end{equation}
At  this  point,  acting  with  $D^{(1)}$  on  $\tilde\Gamma_{\widehat
{A}_{\mu}  \widehat{A}_{\nu}}^{(1)}$  of  (\ref{finBFM}), and  setting
afterwards  $\xi_Q   =1$,  we   immediately  recover  the   result  of
(\ref{sumctBFM}), as announced.

\section{Deformations of the Nielsen Identities}

One of the main advantages of the $D$-formalism is that it allows to obtain
complete control over the  deformations, caused by renormalization, on
the NI, which describe the GFP dependence of
the bare Green functions.

To be specific, let us consider the Abelian Higgs-Kibble model in the
symmetric phase, described by the Lagrangian (see also Appendix A)
\begin{equation}
{\cal L}\ =\ {\cal L}_{\rm I}\: +\: {\cal L}_{\rm GF}\: +\: {\cal
L}_{\rm FP}\; .
\end{equation}
Here, ${\cal L}_{\rm I}$ is the $U(1)$ invariant term
\begin{equation}
{\cal L}_{\rm I}\ =\ -\frac14F^{\mu\nu}F_{\mu\nu}+
\left(D_\mu\Phi\right)^*\left(D^\mu\Phi\right)
-m^2\Phi^*\Phi-\lambda\left(\Phi^*\Phi\right)^2,
\end{equation}
with
\begin{equation}
F_{\mu\nu}=\partial_\mu A_\nu-\partial_\nu A_\mu\;, \qquad
D_\mu=\partial_\mu-igY A_\mu\;,\qquad
\Phi=\frac1{\sqrt2}(H+iG)\;,
\end{equation}
and  the  hypercharge assignment  $Y_\Phi=1$.   The generic gauge-fixing  and
Faddeev-Popov ghost  terms  are given by
\begin{eqnarray}
{\cal L}_{\rm GF}&=&\frac\xi2B^2+BF\;,\nonumber\\
{\cal L}_{\rm FP}&=&-\bar c(s F)\;,
\end{eqnarray}
where  $B$  is  an  auxiliary  non propagating  field  (which  can  be
eliminated through its  equations of motion), $F$ is  the gauge-fixing
function, and  $s$ is the BRS  operator, giving rise  to the following
field transformations
\begin{eqnarray}
sA_\mu=\partial_\mu c, \qquad &sH=-gcG,& \qquad sG=gcH, \nonumber \\
sc = 0, \qquad &s\bar c = B,& \qquad sB=0,
\label{BRS1}
\end{eqnarray}
with $s$ nilpotent. Evidently, ${\cal L}$ is BRS invariant.

Let  us   now promote  the constant  $\xi$  to  a (static)  field, and
introduce an associate anticommuting BRS source $\eta$ (with $\eta^2 =
0$)~\cite{footnote4} through the transformations
\begin{equation}
s\xi=\eta\,, \qquad s\eta=0\; .
\label{BRS2}
\end{equation}
The  BRS invariance  of the  original Lagrangian  ${\cal L}$ is then
lost, unless we add to it the term
\begin{equation}
{\cal L}_{\rm N}\ =\ \frac12\eta\bar c B\;,
\end{equation}
which  couples the  $\eta$  source  to the  other  fields. With  the
Lagrangian term~${\cal L}_{\rm N}$  included, it can be easily checked
that the so-quantized Lagrangian,
\begin{equation}
{\cal L}\ =\ {\cal L}_{\rm I}\: +\: {\cal L}_{\rm GF}\: +\: 
           {\cal  L}_{\rm FP}\: +\: {\cal L}_{\rm N}\; ,
\end{equation}
is  invariant under  the {\em  extended} BRS  ($e$BRS) transformations
given  by (\ref{BRS1})  and  (\ref{BRS2}).  If  we  integrate out  the
auxiliary field $B$, we find
\begin{equation}
B=-\frac1\xi F-\frac12\eta\bar c, \qquad {\cal L}_{\rm GF}=-\frac1{2\xi}F^2,
\qquad {\cal L}_{\rm N}=-\frac1{2\xi}\eta\bar cF.
\label{Bout}
\end{equation}
In  this case,  however, the  $e$BRS transformation  of  the antighost
$\bar c$ is not nilpotent, because one has
\begin{equation}
s^2\bar c\ =\  \frac1\xi\left(-sF+\frac1{2\xi}\eta F\right),
\end{equation}
where  the  term  in  parenthesis  is precisely  the  antighost  field
equation of  motion. Hence,  in this case,  the $e$BRS  algebra closes
on-shell.

The extended Slavnov--Taylor  identities one
obtains from the $e$BRS invariant Lagrangian are precisely the NI.
Specifically, assuming that  the $B$  field has
been  integrated out,  the generating  functional for  connected Green
functions of our model can be written as
\begin{eqnarray}
e^{iZ}&=&\int \![dA_\mu][dc][d\bar c][dH][dG]\, 
\exp\bigg\{i\int\!d^4 x\,\Big[{\cal L}+J_A^\mu A_\mu+J_cc+J_{\bar
c}\bar c +J_HH+J_GG\nonumber\\
&&+ K_H(sH)+K_G(sG)+K_{\bar c}(s\bar
c)\Big]\bigg\}\,,
\end{eqnarray}
where the  $e$BRS sources $K$  (sometimes called the  antifields) have
been  included only  for fields  transforming non  linearly  under the
algebra of  (\ref{BRS1}) and (\ref{BRS2}).  Notice  that the antighost
source $K_{\bar  c}$ must  be included, since  the $B$ field  has been
integrated out  [and therefore $\bar c$  transforms non-linearly under
$e$BRS,  see (\ref{Bout})]. Observe  that, given  a field  $\Phi$, its
corresponding BRS  source $K_\Phi$  obeys opposite statistic,  and has
ghost    charge    $Q_{\mathrm{FP}}(K_\Phi)=-Q_{\mathrm{FP}}(\Phi)-1$;
therefore     one    has    $Q_{\mathrm{FP}}(K_{A_\mu},K_H,K_G,K_{\bar
c})=\{-1,-1,-1,0\}$ (that $K_{\bar{c}}$ has no ghost charge is evident
from the  fact that it couples to  $s\bar{c} = B$ and  the action must
have zero ghost charge).

Thus, eventually the   statement   of   $e$BRS   invariance   
implies   the   following Slavnov--Taylor identity:
\begin{eqnarray}
Z[J_A^\mu,J_c,J_{\bar c},J_H,J_G,K_{\bar c},K_{H},K_{G};\xi,\eta]&&\nonumber\\
&&\hspace{-6cm}=\ 
Z[J_A^\mu,J_c-\omega\partial_\mu J_A^\mu,J_{\bar c},J_H,J_G,K_{\bar
      c}+\omega J_{\bar c},K_{H}+\omega J_H,K_{G}+\omega
    J_G;\xi-\omega\eta,\eta]\;,\qquad 
\end{eqnarray}
which,  after  a Taylor  expansion  in  $\omega$ (with  $\omega^2=0$),
reduces to the (all-order) NI
\begin{equation}
J_A^\mu\partial_\mu\frac{\delta Z}{\delta J_c}+J_{\bar c}\frac{\delta
Z}{\delta K_{\bar c}}+ J_H\frac{\delta Z}{\delta K_H}+J_G\frac{\delta
Z}{\delta K_G}-\eta\partial_\xi Z\ =\ 0\; .
\end{equation}

Denoting  by  $\varphi$  all  the  fields of  our  model,  {\it  i.e.},
$\varphi=\{A_\mu,c,\bar  c,H,G\}$,   one  can  express   the  previous
Slavnov--Taylor   identities  in   terms  of   the   effective  action
$\g[\varphi_{\rm  cl},  K_\varphi;\xi,\eta]$,  given by  the  Legendre
transform
\begin{equation}
\g[\varphi_{\rm cl}, K_\varphi;\xi,\eta]\ =\ Z[J_{\varphi},
K_\varphi;\xi,\eta]\: -\: \int d^4x J_\varphi\varphi_{\rm cl}\; ,
\end{equation}
with  $\varphi_{\rm  cl}=  \delta  Z/\delta J_\varphi$.   Taking  into
account the relations
\begin{equation}
\frac{\delta\g}{\delta\varphi_{\rm cl}}=-J_\varphi\,, \qquad 
\frac{\delta\g}{\delta K_\varphi}=\frac{\delta Z}{\delta K_\varphi}\ ,
\qquad \partial_\xi\g=\partial_\xi Z\;,
\end{equation}
we obtain
\begin{equation}
\frac{\delta\g}{\delta A^\mu_{\rm cl}}\partial_\mu c_{\rm cl}+
\frac{\delta\g}{\delta\bar c_{\rm cl}}\frac{\delta\g}{\delta K_{\bar
c}}+ \frac{\delta\g}{\delta H_{\rm cl}}\frac{\delta\g}{\delta K_H}+
\frac{\delta\g}{\delta G_{\rm cl}}\frac{\delta\g}{\delta
K_G}+\eta\partial_\xi\g\ =\ 0\;.
\end{equation}
Finally,  after  differentiating with  respect  to $\eta$  and setting
$\eta$ afterwards to  zero (with $\eta^2 = 0$),  we obtain  the NI
in its final form
\begin{equation}
  \label{NIbare}
\partial_\xi\g\Big|_{\eta=0}\ =\ 
-\, \partial_\eta \left(\,\frac{\delta\g}{\delta
A^\mu_{\rm cl}}\partial_\mu c_{\rm cl}\: +\: \frac{\delta\g}{\delta\bar
c_{\rm cl}}\frac{\delta\g}{\delta K_{\bar c}}\: +\: \frac{\delta\g}{\delta
H_{\rm cl}}\frac{\delta\g}{\delta K_H}\: +\: \frac{\delta\g}{\delta
G_{\rm cl}}\frac{\delta\g}{\delta K_G}\,\right)\; .
\end{equation}
Introducing the so-called  Slavnov--Taylor operator ${\cal S}_{\g}$, 
the NI~(\ref{NIbare}) is often written down as 
\begin{equation}
  \label{NIST}
\partial_\xi\g\Big|_{\eta=0}\ =\ 
{\cal S}_{\g}\,\Big[\,\partial_\eta \g\, \Big]\; ,
\end{equation} 
where we  have used that $[{\cal  S}_{\g }\,,\, \partial_\eta  ] = 0$.
It is  now easy  to see  that, due to  ghost charge  conservation, the
first term on the RHS of (\ref{NIbare}) becomes relevant only when one
is  calculating the  gauge-dependence of  ghost Green  functions.  For
instance,  it  contributes  a  term  $-\partial_\mu\Gamma_{\bar  c\eta
A_\mu}$ for the NI of the ghost self-energy $\Gamma_{c\bar c}$.

\begin{figure}[t]
\bce
\includegraphics[width=14cm]{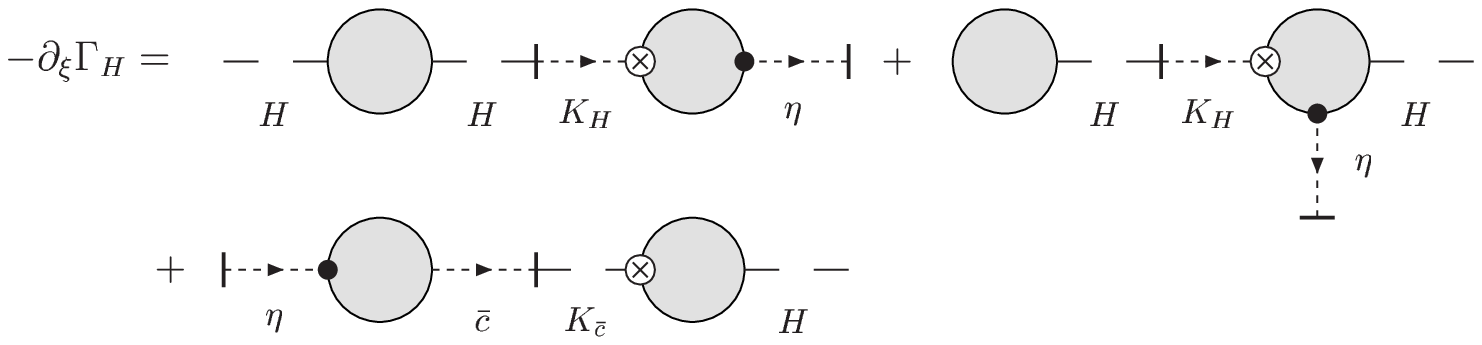}
\caption{\label{NIgraph}   Graphic  representation   of   the  NI
(in  this case  the one for  the tadpole). Notice  that the
arrows indicate the flow of the  ghost charge. The last term is absent
to all orders in the $R_\xi$ gauge.}  
\ece
\end{figure}

A graphical representation  of the NI can be  obtained as follows. The
$\eta$ coupling can be represented by a heavy dot attached to a dotted
line   with  an   arrow  that   indicates  the   flow  of   the  ghost
charge.  Likewise, the  coupling of  the BRS  sources $K_H$  and $K_G$
should by  represented by  a $\otimes$ symbol  attached to  an arrowed
dotted line. On  the other hand, the coupling  of $K_{\bar{c}}$ should
be  represented by  $\otimes$ attached  to  a dashed  line without  an
arrow,  since  $K_{\bar{c}}$ has  no  ghost  charge.  In this  way,  a
consistent graphical picture conveying  information of the flow of the
ghost charge emerges. For example,  the NI for the Higgs tadpole, that
reads (assuming $CP$ conservation)
\begin{equation}
  \label{NItad}
-\partial_\xi\Gamma_H \ \equiv\ -{\cal S}_{\g}\Big[\, 
\partial_\eta \Gamma_H\,\Big]\
 =\ \Gamma_{HH}\Gamma_{\eta
 K_H}+\Gamma_H\Gamma_{\eta K_H H}+\Gamma_{\eta\bar c}\Gamma_{K_{\bar
 c}H}\; ,
\end{equation}
can  be represented  as  in Fig.~\ref{NIgraph}.  It  can be explicitly
checked that the last term on the RHS  of~(\ref{NItad}) is zero to all
orders in perturbation theory in the conventional $R_\xi$ gauge-fixing
scheme.

As already  mentioned in the Introduction,  unlike the Slavnov--Taylor
identities which  are derived from  the BRS invariance of  the theory,
the NI are obtained from  the $e$BRS symmetry; therefore they will not
remain unmodified by the process  of renormalization, but they will be
deformed.   In  what follows,  we  are  going  to determine  a  closed
expression  for the deformation  of the  NI under  renormalization, by
applying the $D$-formalism.

Let us therefore denote by $x$ the parameters of the theory (coupling,
masses, etc.)   {\it except of} the GFP  $\xi$, and  with $\phi_i$ the
fields  $A_\mu, H,$ and $G$,  with renormalization  relations given by
(\ref{genth}).  To  elucidate our points,   we will initially consider
Green functions involving $n$ fields of one type, say $\phi$, and then
we will generalize our results to  Green functions involving different
types of  fields.  Based on the   basic formul\ae{} (\ref{master}) and
(\ref{master2}) of the $D$-formalism, we can now study the response of
the     renormalized     Green     functions    to       a   variation
$\Delta\xi_{\mathrm{R}}$ of  the  renormalized GFP $\xi_{\mathrm{R}}$,
{\it i.e.},
\begin{equation}
\label{var}
\phi^n_{\mathrm R}\Gamma^{\mathrm R}_{\phi^n}(\xr,\xir+\dxir;\mu)-
\phi^n_{\mathrm R}\Gamma^{\mathrm R}_{\phi^n}(\xr,\xir;\mu)\ =\
\phi^n_{\mathrm R}\dxir\partial_{\xir}\Gamma^{\mathrm
R}_{\phi^n}(\xr,\xir;\mu)\ +\ {\cal{O}}(\dxir^2)\, .
\end{equation}
In stating~(\ref{var}), we have assumed  that the renormalized  fields
and parameters $\phi_{\rm R}$ and $x_{\rm R}$ are GFP-independent in a
given renormalization scheme~R.   In fact, this   is the case, if  the
renormalized  parameters are evaluated from renormalization conditions
that are manifestly  GFP independent, {\it e.g.},~from physical  observables,
such as $S$-matrix elements, or from gauge-invariant operators, within
a good    regularization  scheme that  preserves   the Slavnov--Taylor
identities.   In general,  the  GFP  independence of  the renormalized
parameters in a specific scheme can be determined  in two ways: ({\it i})~by
showing  that  the  corresponding  CTs  are GFP-independent;   ({\it ii})~by
comparing   it to   another  gauge-invariant    scheme, such  as   the
$\overline{\rm       MS}$~\cite{MSbar}            or   the   pole-mass
scheme~\cite{Stuart:1991xk,Sirlin:1991fd}.  An exception to the  above
are the renormalized field  $\phi_{\rm  R}$ and  its VEV $v_{\rm  R}$.
These   parameters are     GFP-independent,   but  their    respective
renormalizations    $Z_\phi^{\frac12}$  and    $Z_v$ are   in   general GFP-{\em
dependent}, even   within   the $\overline{\rm  MS}$    scheme.  Thus,
although we will initially assume that  $\phi_{\rm R}$ and $x_{\rm R}$
are  GFP-independent, we will  also comment on  the modifications that
should be considered, if $x_{\rm R}$ were depending on $\xi_{\rm R}$.

Let us concentrate on  the LHS of  (\ref{var}), and more  precisely on
its first term. {}From (\ref{Gren}), we will have that
\begin{eqnarray}
  \label{var2}
\phi^n_{\mathrm R}\Gamma^{\mathrm R}_{\phi^n}(\xr,\xir+\dxir;\mu) &=&
\Big(\phi_{\mathrm R}+\delta\phi(\xir+\dxir)\Big)^n \nonumber \\
&&\times\,\Gamma_{\phi^n}\Big(\xr+\delta
x(\xir+\dxir),\xir+\dxir+\delta\xi(\xir+\dxir);\mu,\epsilon\Big)\nonumber\\ 
&=& \Big(\phi_{\mathrm R}+\delta\phi(\xir)\Big)^n\;
\Gamma_{\phi^n}\Big( \xr+\delta x(\xir), 
\xir+\delta\xi(\xir);\mu,\epsilon\Big)\nonumber \\ 
&&+\,\dxir
\Big( \phi_{\mathrm R}+\delta\phi(\xir)\Big)^n\Big[\, n \Big(
\phi_{\mathrm R}+\delta\phi(\xir)\Big)^{-1}
\Big(\partial_{\xir}\delta\phi(\xir)\Big)\nonumber\\
&&+\,\Big(\partial_{\xir}\delta x(\xir)\Big)\partial_{\xr+\delta x}\:
+\: \Big( 1+\partial_{\xir}\delta \xi(\xir)\Big) \partial_{\xir+\delta
\xi}\Big]\nonumber\\ 
&&\times\, \Gamma_{\phi^n}\Big(\xr+\delta
x(\xir),\xir+\delta\xi(\xir);\mu,\epsilon\Big)\ +\ {\cal{O}}(\dxir^2)\;.
\end{eqnarray}
Substituting   for  the   shifts   their  expressions   in  terms   of
renormalization  constants [cf.~(\ref{CTshift})],  and  then inserting
the  result  back  into  (\ref{var2}),   we  obtain  for  the  RHS  of
(\ref{var}) that
\begin{eqnarray}
\phi^n_{\mathrm R}\partial_{\xir}\Gamma^{\mathrm
R}_{\phi^n}(\xr,\xir;\mu) \!&=&\!
\phi^n\, \Big[\partial_{\xir}(Z_\xi\xir)\Big]\,\partial_\xi
\Gamma_{\phi^n}(x,\xi;\mu,\epsilon)\, +\,
Z_{\phi}^{\frac n2}\phi^n_{\mathrm R}\Big(\partial_{\xir} Z_x\Big)\, 
x_{\rm R}\,\partial_x \Gamma_{\phi^n}(x,\xi;\mu,\epsilon) \nonumber\\ 
&&+\, n\phi^n_{\mathrm R}\, \Big( \partial_{\xir}\ln
Z_{\phi}^{\frac12}\Big)\, \Gamma^{\mathrm R}_{\phi^n}(\xr,\xir;\mu)\; ,
\end{eqnarray}
which is equivalent to the identity 
\begin{eqnarray}
  \label{NIxi}
\partial_{\xir}\Gamma^{\mathrm R}_{\phi^n}(\xr,\xir;\mu) &=&
Z_{\phi}^{\frac n2}\Big[\partial_{\xir}(Z_\xi\xir)\Big]\partial_\xi
\Gamma_{\phi^n}(x,\xi;\mu,\epsilon)\, +\, Z_{\phi}^{\frac
n2}\Big( \partial_{\xir} Z_x\Big) x_{\rm R}
\partial_x\Gamma_{\phi^n}(x,\xi;\mu,\epsilon)\nonumber\\ 
&&+\, n\,\Big(\partial_{\xir}\ln Z_{\phi}^{\frac12}\Big)\,\Gamma^{\mathrm
R}_{\phi^n}(\xr,\xir;\mu)\; .
\end{eqnarray}
Finally,  including   $\xi$  in  the  parameters  $x$,   we  may  cast
(\ref{NIxi}) into the slightly more compact form:
\begin{eqnarray}
  \label{NIren} 
\partial_{\xir}\Gamma^{\mathrm R}_{\phi^n}(\xr;\mu)&=&
Z_{\phi}^{\frac  n2}Z_\xi\partial_\xi\Gamma_{\phi^n}(x;\mu,\epsilon)\:
+\:   Z_{\phi}^{\frac  n2}\,   \Big(  \partial_{\xir} Z_x\Big)\,
x_{\rm R}\,\partial_x      \Gamma_{\phi^n}(x;\mu,\epsilon)\nonumber\\   
&&+\,n\,\Big(\partial_{\xir}\ln  Z_{\phi}^{\frac12}\Big)\,     
\Gamma^{\mathrm R}_{\phi^n}(\xr;\mu)\; .
\end{eqnarray}
A similar line of arguments can be followed, if $x_{\rm R}$ depends on
$\xi_{\rm R}$.  In this case, the factor $(\partial_{\xir} Z_x) x_{\rm
R}$ contained in the second term on the RHS of~(\ref{NIren}) should be
replaced with $\partial_{\xir}( Z_x x_{\rm R}  )$, for $x_{\rm R} \neq
\xi_{\rm R}$.

We  may  now  rely     on  the  $D$-formalism  to   express   the  RHS
of~(\ref{NIren}) {\em entirely}  in terms of renormalized  parameters.
In this way, we obtain the deformed NI
\begin{eqnarray}
  \label{dNI}
\partial_{\xir}\Gamma^{\mathrm R}_{\phi^n}(\xr;\mu)&=& 
Z_{\phi}^{\frac
n2}Z_\xi\:\Big<\, e^D\, {\cal S}_{\g } \Big[\, \partial_\eta
\Gamma_{\phi^n}(x_{\rm R};\mu,\eta,\epsilon)\,\Big]\,\Big>\\ 
&&+\, Z_{\phi}^{\frac  n2}\,   \Big(  \partial_{\xir}Z_x\Big)\, x_{\rm R}\,\:
\Big<\, e^D\, \partial_{x_{\rm R}}      
\Gamma_{\phi^n}(x_{\rm R};\mu,\epsilon)\,\Big>\: +\: 
n\, \Big(\partial_{\xir}\ln Z_{\phi}^{\frac12}\Big)\, \Gamma^{\mathrm
R}_{\phi^n}(\xr;\mu)\; .\nonumber
\end{eqnarray}
Notice that, up to an overall constant, only the first term on the RHS
of~(\ref{dNI})  can  be  related  to  the  undeformed,  unrenormalized
NI~(\ref{NIST}), where the 1PI  Green functions involved are evaluated
with renormalized parameters.  The appearance  of the other terms is a
consequence of the renormalization process.  We emphasize that, unlike
earlier formal  treatments~\cite{Gambino:1999ai}, Equation (\ref{dNI})
furnishes for  the first  time the precise  closed and,  in principle,
calculable expression for the deformations of the NI.

It is interesting to remark that even the action of the operator $e^D$
present  in the first  term on  the RHS  of~(\ref{dNI}) gives  rise to
further deformations.  To make this last point more explicit, we first
observe that the action of the $D$ operator in~(\ref{dNI}) is actually
equivalent  to  the  action  of a  reduced  operator  $\widetilde{D}$,
without the field derivatives, {\it i.e.},~$\widetilde{D} = D - \sum_i
\delta\phi_i\,\partial/\partial\phi_{i{\rm  R}}$.  Making use  of this
last    observation,    we    decompose   $e^D$,    or    equivalently
$e^{\widetilde{D}}$, as $e^{\widetilde{D}}  = 1 + (e^{\widetilde{D}} -
1)$.   Subsequently, noticing that  the operators  $\widetilde{D}$ and
${\cal S}_{\g}$ commute, {\it i.e.},~$[ \widetilde{D}\,, {\cal S}_{\g}
] = 0$, the first term  on the RHS of~(\ref{dNI}) can be rewritten, up
to the multiplicative factors $Z_{\phi}^{\frac n2}Z_\xi$, as
\begin{eqnarray}
 \label{Drest}
\Big<\, e^D\, {\cal S}_{\g } 
\Big[\, \partial_\eta \Gamma_{\phi^n}(x_{\rm R};\mu,\eta,\epsilon)\,
\Big]\,\Big> &=& \Big<\, {\cal S}_{\g } 
\Big[\, \partial_\eta \Gamma_{\phi^n}(x_{\rm R};\mu,\eta,\epsilon)\,
\Big]\,\Big>\nonumber\\ 
&&+\: \Big<\, {\cal S}_{\g } \Big[\, (e^{\widetilde{D}} - 1)\, 
\partial_\eta \Gamma_{\phi^n}(x_{\rm R};\mu,\eta,\epsilon)\,
\Big]\,\Big>\; .
\end{eqnarray}
Thus, only the first term can be expressed in terms of the undeformed,
unrenormalized  NI of (\ref{NIST}), whereas the second one
is an additive deformation of the NI that results in from a
BRS variation of another   function~\cite{Gambino:1999ai}.  

The  generalization   of~(\ref{dNI})  to  Green   functions  involving
different  types of  fields  is straightforward~\cite{footnote5},  and
reads
\begin{eqnarray}
  \label{NIgeneral}                     
\partial_{\xir}\Gamma^{\mathrm R}_{\prod_i\phi^{n_i}_i}(x_{\mathrm
R};\mu) &=& \prod_i Z_{\phi_i}^{\frac{n_i}2}Z_\xi\, \Big<\, e^D\,
{\cal S}_{\g } \Big[\, \partial_\eta\, \Gamma_{\prod_i\phi^{n_i}_i}(x_{\rm
R};\mu,\eta,\epsilon)\, \Big]\, \Big>\nonumber\\
&& +\: \prod_iZ_{\phi_i}^{\frac{n_i}2} \sum_x \Big( \partial_{\xir}
Z_x\Big)\,  x_{\rm R}\: \Big<\, e^D\, \partial_{x_{\rm R}}
\Gamma_{\prod_i\phi^{n_i}_i}(x_{\rm R};\mu,\epsilon)\,\Big>\\ 
&&+\: \sum_i n_i\Big(\partial_{\xir}\ln Z_{\phi_i}^{\frac12}\Big)
\Gamma^{\mathrm R}_{\prod_i\phi^{n_i}_i}(x_{\mathrm R};\mu)\; .\nonumber
\end{eqnarray}

In order to get a feel on the structure of Equation (\ref{NIgeneral}),
let  us   apply  it  to  the  lowest   non-trivial  order.   Expanding
consistently, according to the  $D$-formalism, we obtain the following
one-loop result:
\begin{eqnarray}
\partial_{\xir}\Gamma^{\mathrm R\,(1)}_{\prod_i\phi^{n_i}_i}(x_{\mathrm
  R};\mu) &=&\partial_{\xir}\Gamma^{(1)}_{\prod_i\phi^{n_i}_i}(x_{\rm
  R};\mu,\epsilon) 
+\sum_i\frac{n_i}{2}\left(\partial_{\xir}Z^{(1)}_{\phi_i}\right)\Gamma^{\mathrm
  R\, (0)}_{\prod_i\phi^{n_i}_i}(x_{\mathrm R})\nonumber\\ 
&&+\Big(\sumi\frac{n_i}{2}Z^{(1)}_{\phi_i}+Z^{(1)}_\xi+\sum_xZ^{(1)}_xx_{\rm
  R}\partial_{x_{\rm R}}\Big)\partial_{\xir}  
\Gamma^{(0)}_{\prod_i\phi^{n_i}_i}(x_{\rm R})\nonumber\\
  &&+\sum_x\left(\partial_{\xir}Z^{(1)}_{x}\right)x_{\rm
  R}\partial_{x_{\rm R}}  
\Gamma^{(0)}_{\prod_i\phi^{n_i}_i}(x_{\rm R})\; ,
\label{1ldefNI}
\end{eqnarray}
where the  first term is the  one-loop NI of  (\ref{NIbare}), with the
simple replacement of bare parameters by renormalized ones.

A  further   simplification of   the    formulae above occurs   in   a
gauge-invariant    renormalization   scheme,         such   as     the
$\overline{\mathrm{MS}}$ scheme.  In this case, all terms proportional
to    $\partial_{\xir} Z_x$, for  which     $x$  is related to    a
gauge-invariant operator ({\it  e.g.},~a gauge-coupling  constant or a
gauge-invariant   mass parameter),  will  drop   out  from the  RHS of
(\ref{NIgeneral}), as   their multiplicative renormalization constants
will be $\xi_{\rm R}$-independent.

\section{The Abelian Higgs model in the HLET-preserving gauges}

In this section  we will discuss the Abelian  Higgs model quantized in
the type  of gauges that  preserve the Higgs-boson Low  Energy Theorem
(HLET).  The Lagrangian defining the model is given by
\begin{equation}
  \label{LU1}
{\cal L}\ =\ {\cal L}_{\rm I}\: +\: {\cal L}_{\rm GF}\: +\: 
{\cal L}_{\rm FP}\; .
\end{equation}
Here ${\cal L}_{\rm I}$ is the gauge-invariant part of the Lagrangian
\begin{eqnarray}
  \label{LI}
{\cal L}_{\rm I}&=&-\frac14F^{\mu\nu}F_{\mu\nu}+
\left(D_\mu\Phi\right)^*\left(D^\mu\Phi\right)
-m^2\Phi^*\Phi-\lambda\left(\Phi^*\Phi\right)^2\nonumber\\
&&+\, \sumi\bar f_i^L(i\gamma^\mu D_\mu)f_i^L+\sumi\bar
f^{R}_i(i\gamma^\mu \partial_\mu)f^{R}_i \nonumber\\ 
&&-\,\sqrt2h_1\bar f^L_1\Phi f^{R}_1-\sqrt2h_2\bar f^L_2\Phi^* f^{R}_2 
-\sqrt2 h_1\bar f^{R}_1\Phi^* f^L_1-\sqrt2h_2\bar f^{R}_2\Phi f^L_2\ .
\end{eqnarray}
In (\ref{LI}), $F_{\mu\nu}=\partial_\mu A_\nu-\partial_\nu A_\mu$ is
the $U(1)_Y$ Field strength tensor,  $D_\mu=\partial_\mu-igY A_\mu$
is the corresponding covariant derivative, and
\begin{equation}
\Phi\ =\ \frac1{\sqrt2}\; \Big(\, v\, +\, H\, +\, iG \Big) 
\end{equation}
is the complex Abelian Higgs  field, composed from the two real fields
$H$ and $G$, where $\langle \Phi \rangle = v/\sqrt{2}$ is its VEV that
signifies  the   spontaneous  symmetry  breaking   of  $U(1)_Y$.   The
hypercharge  quantum  numbers of  the  different  fields are  assigned
according  to $Y_\Phi=1$,  $Y_1^L=-Y_2^L=1$,  and $Y_1^{R}=Y_2^{R}=0$.
For simplicity, we finally assume  that the Yukawa couplings $h_i$ are
real.

For the gauge-fixing term  ${\cal L}_{\rm GF}$ in~(\ref{LU1}), we will
adopt the  one introduced in \cite{Kastening:1993zn,Pilaftsis:1997fe},
{\it i.e.},
\begin{eqnarray}
{\cal L}_{\rm GF}&=&-\frac\sigma{2\xi}[\partial_\mu A^\mu+\xi g\,{\rm
    Im}(\Phi^2)]^2\nonumber \\ 
&=&-\frac\sigma{2\xi}\Big[\partial_\mu A^\mu\: +\: \xi g(v+H)\,G\,\Big]^2\,,
\label{LGF}
\end{eqnarray} 
which in turn induces the Faddeev--Popov ghost term
\begin{eqnarray} {\cal L}_{\rm
FP}&=&-\bar c\left[\partial^2+2g^2\xi\,{\rm
Re}(\Phi^2)\right]c\nonumber\\ &=&-\bar c\,
\Big\{ \partial^2 + g^2\xi\Big[ (v+H)^2-G^2\Big]\Big\}\,c\; .  
\end{eqnarray}
Notice  the  presence  of  the  two GFPs,  $\xi$  and  $\sigma$.   The
gauge-fixing  scheme   described  by  ${\cal  L}_{\rm   GF}$  and  the
BRS-induced ghost term ${\cal L}_{\rm  FP}$ (which will be referred to
as  the  $\overline{R_\xi}$  scheme),  in addition  to  the  technical
advantages mentioned in \cite{Kastening:1993zn}, belong to the special
class     of      schemes     which     respect      the     so-called
HLET~\cite{Ellis:1975ap,Shifman:1978zn,
Shifman:1979eb,Vainshtein:1980ea,Dawson:1989yh}      for     off-shell
unrenormalized      Green      functions      beyond     the      tree
level~\cite{Pilaftsis:1997fe}  (see also  our discussion  below).  The
reason is that the complete Lagrangian, including the gauge-fixing and
ghost    sectors,    is     invariant    under    the    translational
transformation~\cite{Pilaftsis:1997fe}:
\begin{equation}
  \label{trans}
H\ \to\  H\: +\: a\,, \qquad  v\ \to\   v\: -\: a\, .
\end{equation}
For example, the conventional  ${\rm R_{\xi}}$ gauge, described by the
gauge-fixing term,
\begin{equation}
{\cal L}_{\rm R_{\xi}}\ =\ \frac{1}{2\xi}\; \Big[\,\partial_\mu
A^\mu\: +\: \frac{1}{\sqrt{2}}\, \xi g  v\,{\rm Im}(\Phi)\,\Big]^2\, ,
\end{equation}
violates this translational symmetry~(\ref{trans}) due to its explicit
dependence on $v$.  As a consequence of this violation, the HLET given
below by (\ref{HLETGreen}) is no longer valid beyond the tree level.

The  translational symmetry~(\ref{trans})  is sufficient  to  show the
validity of the HLET to all  orders. As a result of this symmetry, the
entire effective action $\g[H,v]$  (where we suppress all other fields
and parameters) satisfies the identity
\begin{equation}
\g[H,v]\ =\ \g[H+a,v-a],
\end{equation}
which immediately implies the translational Ward identity 
\begin{equation}
\frac{\delta\g}{\delta H}\ =\ \frac{\partial\g}{\partial v}\ .
\label{twi}
\end{equation}
In  general, upon $n$  functional differentiations with respect to the
field $H$, we obtain the HLET for off-shell (unrenormalized) $n$-point
1PI Green functions $\Gamma_{H^n}$:
\begin{equation}
  \label{HLETGreen}
\Gamma_{H^{n+1}}\ =\ \frac{\partial\Gamma_{H^n}}{\partial v}\ .
\end{equation}
In the above formula,  the Higgs-boson insertion in $\Gamma_{H^{n+1}}$
is evaluated at  zero momentum. This result should  be contrasted with
the one obtained in the $R_\xi$  gauge, in which the HLET described by
the   relation~(\ref{HLETGreen})   will   be   grossly   violated   by
gauge-mediated quantum effects~\cite{Pilaftsis:1997fe}.

The translational identity (\ref{twi}) has an immediate consequence on
the  way the  Higgs  VEV is  renormalized.   The most  general way  of
renormalizing $v$ is given by \cite{Bohm:1986rj,Chankowski:1991md}
\begin{equation} 
v\ =\ Z_v v_{\mathrm R}\ =\ Z_H^\frac{1}{2}(v_{\mathrm R} + \Delta v) \;.
\end{equation}
It  should be  remembered  that  $\Delta v$  differs  from $\delta  v$
defined previously in~(\ref{CTshift}), since~$\delta  v = (Z_v - 1 )\,
v_{\rm R}$.  The quantity $\Delta v$  may be split into two parts: one
part  that   contains  the  divergent   contribution  proportional  to
$1/\epsilon$, to  be denoted by  $\Delta v|_{\rm div}$, and  a finite,
renormalization-scheme  dependent  piece,  to  be denoted  by  $\Delta
v|_{\rm fin}$,  {\it i.e.}, $\Delta v  = \Delta v|_{\rm  div} + \Delta
v|_{\rm fin}$.  The  crucial point is that if the  HLET is exact, then
one  must have  that  $\Delta v|_{\rm  div}  =0$.  To  prove this,  we
observe that the Higgs  tadpole $\Gamma_H$ and the effective potential
$\Gamma$,   by   virtue   of   (\ref{twi}),   satisfy   the   equality
$\Gamma_H=\partial_v\Gamma$.      Given     that    $\Gamma_H^{\mathrm
R}=Z_H^{\frac12}\Gamma_H$, we find
\begin{equation}
\Gamma_H^{\mathrm R}\ =\ Z_H^{\frac12}\partial_v\Gamma\ =\ 
Z_H^{\frac12}Z_v^{-1}\partial_{v_{\mathrm R}}\Gamma\ =\ 
Z_H^{\frac12}Z_v^{-1}\partial_{v_{\mathrm R}}\Gamma^{\mathrm R}\; ,
\end{equation}
where   in   the   last   step    we   have   used   the   fact   that
$\Gamma(x;\mu,\epsilon)  =   \Gamma^{\mathrm  R}(x_{\mathrm  R};\mu)$.
Therefore we get the condition
\begin{equation}
Z_H^{\frac12}Z_v^{-1}\ =\ c_{\rm fin}\; ,
\end{equation}
where $c_{\rm fin}$ is an UV finite constant.  In perturbation theory,
this finite constant may be decomposed as $c_{\rm fin}=1+\delta c_{\rm
fin}$, where $\delta c_{\rm fin}$ is a higher-order scheme dependence.
In the  $\overline{\rm MS}$ scheme, the multiplicative renormalization
constants have no UV finite pieces,  so that one would get unavoidably
$c_{\rm  fin}=1$,  with   $\delta   c_{\rm  fin}=0$,  and    therefore
$Z_H^{\frac12} = Z_v$, which is equivalent to $\Delta v|_{\rm div} =0$
to   all  orders,  with   $v=Z_H^{\frac12}v_{\mathrm R}$.   In   other
renormalization schemes,  one  needs to  impose that (\ref{HLETGreen})
holds true after renormalization,  so that again $c_{\rm fin}=1$; this
can be done without  intrinsic inconsistencies in the  HLET preserving
gauges,  by imposing that $Z_H^{\frac12}=Z_v$,  even for the UV finite
pieces.

In the remainder of the section, we will present the Lagrangian of the
Abelian Higgs model,  after the SSB of $U(1)_Y$.   This will enable us
to set up the stage for the two-loop calculations related to the issue
of gauge  dependence of  $\tan\beta$, which will  be discussed  in the
next section.  The full Lagrangian of the $U(1)_Y$ model after SSB may
be written down as a sum of four terms:
\begin{eqnarray}
  \label{LHLET}
{\cal L}_0 &=&-\frac14\lambda v^4-\frac12m^2 v^2\;,\nonumber\\ {\cal
L}_1 &=&-(\lambda v^3+m^2 v)H\;,\nonumber\\ 
{\cal L}^\sigma_2 &=&
\frac{1}{2}(\partial_\mu H)^2-\frac{1}{2}(3\lambda v^2+m^2)H^2
+\frac{1}{2}(\partial_\mu G)^2 -\frac{1}{2}[(\lambda+\sigma\xi g^2)
v^2+m^2]G^2\nonumber\\
&&-\,\frac{1}{4}F_{\mu\nu}F^{\mu\nu}-\frac{\sigma}{2\xi}(\partial_\mu
A^\mu)^2 +\frac{1}{2}g^2 v^2A_\mu A^\mu-(\sigma-1)g v G(\partial_\mu
A^\mu) \nonumber\\ 
&&+\,\partial_\mu\bar c\partial^\mu c -\xi g^2
v^2\bar c c+\sumi\bar f_i(i\gamma^\mu\partial_\mu-h_i v)f_i\;,\nonumber\\ 
{\cal L}^\sigma_{3,4} &=&-\frac{1}{4}\lambda H^4-\lambda v H^3
-\frac{1}{2}(\lambda+\sigma\xi g^2)H^2G^2 -(\lambda+\sigma\xi g^2) v
HG^2 -\frac{1}{4}\lambda G^4\nonumber\\ 
&&+\, g^2 v HA_\mu A^\mu
+\frac{1}{2}g^2H^2A_\mu A^\mu +\frac{1}{2}g^2G^2A_\mu A^\mu
+(\sigma+1)gGA_\mu(\partial^\mu H)\nonumber\\
&&+\, (\sigma-1)gHA_\mu(\partial^\mu G) -2\xi g^2 v H\bar cc -\xi
g^2H^2\bar cc +\xi g^2G^2\bar cc \nonumber\\ 
&&+\, g A^\mu\sumi\bar f_i
Y^L_i\gamma_\mu P_L f_i -\sumi h_i\bar f_iHf_i-\sumi Y^L_ih_i\bar
f_i\gamma_5 Gf_i\;,
\end{eqnarray} 
where  all the  fields and  parameters are  {\it bare}.   Observe that
$\sigma=1$  is the  only value  that  avoids the  appearance of  mixed
propagators  between the would-be  Goldstone boson  $G$ and  the gauge
boson $A_\mu$.   We will therefore  renormalize the model  by imposing
the latter  condition on the  renormalized GFP $\sigma_{\rm  R}$, {\it
i.e.}, $\sigma_{\mathrm R}=1$. This condition becomes rather subtle at
two  loops, and  especially when  the  $D$ operator  is applied  after
integration,  in which  case one  needs to  keep the  (one  loop) full
dependence on $\sigma$ in all  the quantities under study. However, as
we  already  mentioned, our  calculational  task  may considerably  be
simplified if  the $D$-formalism is applied before  the integration of
the loop momenta.

In the  HLET-preserving gauges, the propagators take  on the following
form:
\begin{eqnarray}
D_H(k)\ =\ \frac i{k^2-m^2_H}\ ,&\qquad& \
D_G(k)\ =\ \frac i{k^2-m^2_G}\ ,\nonumber\\
D_c(k)\ =\ \frac i{k^2-m^2_c}\ ,\ 
&\qquad&\Delta_{\mu\nu}(k)\ =\ \frac{i}{k^2-m^2_A}
\left[\,-g_{\mu\nu} + \left(1-\frac\xi\sigma\right)\frac{k_\mu
k_\nu}{k^2-\frac1\sigma m^2_c}\right]\ , \nonumber\\ 
S_i(k)\ =\ \frac i{\kslush-m_i}\ ,\quad&&
\end{eqnarray}
where the particle masses are related to the independent parameters of
the theory through
\begin{eqnarray}
m^2_A &=& g^2 v^2\,, \qquad
m^2_H\ =\ 3\lambda v^2+m^2\,, \qquad
m^2_G\ =\ (\lambda+\sigma\xi g^2) v^2+m^2\,, \nonumber\\
m^2_c &=& \xi m^2_A\,, \qquad \, m_i\ =\ h_i v\; .
\end{eqnarray}
The   complete   set   of   Feynman   rules  is   finally   given   in
Fig.~\ref{feynmanrules} of the Appendix.

\section{Gauge dependence of {\boldmath $\tan\beta$} at two loops}

In models  with two elementary Higgs  bosons, $H_1$ and  $H_2$, one of
the fundamental  parameters is the  ratio of their  vacuum expectation
values,         $v_1$          and         $v_2$,         respectively
\cite{Gunion:1989we,Haber:1978jt}.    In   particular,  the   quantity
usually denoted as $\tan\beta$, is defined at tree-level as $\tan\beta
= v_2/v_1$. When quantum corrections are included $\tan\beta$ develops
a non-trivial dependence on the renormalization mass $\mu$, as well as
on the unphysical  GFP. Given that $\tan\beta$ is  extensively used in
parametrizing new  physics effects in  many popular extensions  of the
Standard Model, such as two-Higgs models and almost all supersymmetric
versions,   this   type   of   gauge-dependence  is   an   undesirable
feature. Various  studies have therefore addressed  the question under
which  conditions  a   gauge-independent  definition  of  the  running
$\tan\beta$ could become possible \cite{Yamada:2001ck,Freitas:2002um}.

In  studying these  issues, there  appears  to be  a subtle  interplay
between being  able to set $\Delta  v|_{div} =0$ and  showing that the
difference $\gamma_{H_1}-\gamma_{H_2}$ of  the anomalous dimensions is
independent of the GFP. In this section we will explore in detail this
connection, and  demonstrate that, contrary to what  one might naively
have expected, it is  not possible to establish the gauge-independence
of  $\gamma_{H_1}-\gamma_{H_2}$, at  least not  within  a conventional
field-theoretic framework.

The   basic   observation   which   suggests  a   link   between   the
gauge-independence of $\tan\beta$ and  $\Delta v|_{\rm div} =0$ is the
following.   If   $\Delta  v_i|_{\rm  div}  =0$,   for  $i=1,2$,  then
$v_i=Z_{H_i}^\frac{1}{2}  v_{i\mathrm R}$,  and  therefore $\tan\beta$
renormalizes as
\begin{equation}
\tan\beta\ \equiv\ \frac{v_2}{v_1}\ =\ 
\frac{Z_{H_2}^\frac{1}{2}}{Z_{H_1}^\frac{1}{2}}\
\frac{v_{2\mathrm R}}{v_{1\mathrm R}}\ =\ 
\frac{Z_{H_2}^\frac{1}{2}}{Z_{H_1}^\frac{1}{2}}\
\tan\beta_{\mathrm R}\; . 
\label{tanbren}
\end{equation}
The renormalization group equation for $\tan\beta_{\mathrm R}$ is thus
given by
\begin{equation}
\frac{d\tan\beta_{\mathrm R}}{dt}\ =\ \Big(\gamma_{H_2}\: -\: 
\gamma_{H_1}\Big)\, \tan\beta_{\mathrm R} \;, 
\label{tanbrg}
\end{equation} 
where $t= \ln\mu$, and  $\gamma_{H_i} = -\frac{1}{2} \frac{d}{dt} (\ln
Z_{H_i})$ is the anomalous dimension  of the Higgs field $H_i$.  If at
this point  one could  show that,  {\it in the  class of  gauges where
$\Delta       v_i|_{\rm      div}      =0$},       the      difference
$(\gamma_{H_2}-\gamma_{H_1})$  is GFP-independent,  one  would have  a
solution  to the  problem.   The  crucial point  in  this argument  is
precisely   that   the  two   conditions   need   be  satisfied   {\it
simultaneously}.    Indeed,  having   a   gauge-fixing  scheme   where
$(\gamma_{H_2}-\gamma_{H_1})$  is GFP-independent  does not  imply, by
virtue of (\ref{tanbrg}),  the GFP-independence of $\tan\beta_{\mathrm
R}$, {\it unless} one could  demonstrate that, within the same scheme,
one is  also able  to set  $\Delta v_i|_{\rm div}  =0$; if  the latter
condition cannot  be enforced the renormalization  group equation that
$\tan\beta_{\mathrm   R}$   satisfies    is   simply   not   that   of
(\ref{tanbrg}),   since  the   starting  assumption   is   not  valid.
Satisfying  both conditions  simultaneously is  far from  trivial.  In
fact, as we  will see in detail, in the context  of both the $R_{\xi}$
and  the   HLET-preserving  $\overline{R_{\xi}}$  gauges,   these  two
conditions {\it cannot} be  simultaneously met, for different reasons.
Specifically,  in the $R_\xi$  gauges each  of the  two-loop anomalous
dimension $\gamma_{H_i}$,  consists of two  pieces: ({\it i})  a gauge
dependent ${\cal O}(g^4)$ polynomial, common to both, and ({\it ii}) a
gauge independent ${\cal O}(h^2g^2)$, which is different for $H_1$ and
$H_2$.  Thus, in  taking the difference $(\gamma_{H_2}-\gamma_{H_1})$,
one    finds   a    GFP-independent   answer    for    this   quantity
\cite{Yamada:2001ck}. However,  since the $R_{\xi}$ gauges  are not of
the HLET-preserving  type, one cannot set $\Delta  v_i|_{\rm div} =0$,
and  therefore  (\ref{tanbrg})  receives additional  (gauge-dependent)
contributions.   On  the other  hand,  the $\overline{R_{\xi}}$  gauge
preserves the HLET by construction,  and one may set $\Delta v_i|_{\rm
div} =0$,  thus enforcing the validity of  (\ref{tanbrg}); however, as
we  will see  in  detail  in what  follows,  the two-loop  calculation
reveals that, in these latter gauges, $(\gamma_{H_2}-\gamma_{H_1})$ is
in fact GFP-dependent.

In order  to demonstrate this,  it is actually sufficient  to consider
the Abelian-Higgs  model, despite the  fact that it contains  only one
Higgs   field.   The   rationale   is   that   in   the   context   of
$\overline{R_{\xi}}$ gauges the ${\cal O}(h^2g^2)$ contributing to the
Higgs   boson    anomalous   dimension   turns   out    to   be   {\it
GFP-dependent}. Therefore,  given that,  in general, each  Higgs boson
couples  differently to  the fermions  ({\it i.e.},  the corresponding
Yukawa  couplings are independent  parameters), even  if there  were a
second Higgs boson, this gauge  dependence could not in general cancel
in the difference $(\gamma_{H_2}-\gamma_{H_1})$.

In the rest of this section  we will prove the gauge dependence of the
${\cal O}(h^2g^2)$ contributions to $\gamma_H^{(2)}$.  To that end, we
will  employ two  independent, but  complementary, approaches.  In the
first  approach  we will  exploit  the validity  of  the  HLET in  the
$\overline{R_{\xi}}$  gauges   in  order  to   eventually  obtain  the
(non-vanishing) first derivative of  $Z^{(2)}_H$ with respect to $\xi$
from  the   two-loop  effective  potential,   {\it  without}  actually
computing  Higgs-boson  self-energies.   Second,  we  will  explicitly
compute  $Z^{(2)}_H$ from the  corresponding two-loop  diagrams. Since
for determining  $Z^{(2)}_H$ one needs to consider  only the divergent
terms proportional to the external momentum $q^2$, whereas terms whose
dimensionality   is  saturated   by  masses   do  not   contribute  to
$Z^{(2)}_H$, one may carry out the calculation in the symmetric phase,
when $v  = 0$.   In both approaches  we will employ  the $D$-formalism
developed in the  previous sections in order to  enforce the necessary
cancellations  of   the  overlapping  divergences,   without  explicit
reference to CT diagrams.

Let us first write down how the relevant Green functions renormalize.
From the second equation of (\ref{exp}), one finds the following
relations 
\begin{eqnarray} 
  \label{renor}
i\left(Z_H Z_{m^2} m^2_{\mathrm
R}+Z^2_HZ_\lambda\lambda_{\mathrm R}v_{\mathrm R}^2\right)^{(2)}
v_{\mathrm R}&=&\left.\left\{\overline{D^{(1)}\Gamma^{(1)}_H(x_{\rm
R};\mu,\epsilon)}+ 
\overline{{\Gamma}^{(2)}_H(x_{\rm
R};\mu,\epsilon)}\right\}\right|_{\sigma_{\rm R}=1}, \nonumber\\ 
i\left(Z_H Z_{m^2} m^2_{\mathrm R}+3Z^2_HZ_\lambda\lambda_{\mathrm
R}v_{\mathrm R}^2\right)^{(2)}&=&\left.\left\{
\overline{D^{(1)}\Gamma^{(1)}_{H^2}(x_{\rm R};\mu,\epsilon)}+
\overline{\Gamma^{(2)}_{H^2}(x_{\rm
R};\mu,\epsilon)}\right\}\right|_{\sigma_{\rm R}=1}, \nonumber \\ 
6i\left(Z^2_HZ_\lambda\right)^{(2)}\lambda_{\mathrm R}v_{\mathrm R}
&=&\left.\left\{\overline{ D^{(1)}\Gamma^{(1)}_{H^3}(x_{\rm R};\mu,\epsilon)}+
\overline{\Gamma^{(2)}_{H^3}(x_{\rm
R};\mu,\epsilon)}\right\}\right|_{\sigma_{\rm R}=1},\nonumber \\ 
6i\left(Z^2_HZ_\lambda\right)^{(2)}\lambda_{\mathrm R}
&=&\left.\left\{ \overline{D^{(1)}\Gamma^{(1)}_{H^4}(x_{\rm R};\mu,\epsilon)}+
\overline{\Gamma^{(2)}_{H^4}(x_{\rm
R};\mu,\epsilon)}\right\}\right|_{\sigma_{\rm R}=1}\; , 
\end{eqnarray} 
where $x_{\rm R}$ collectively denotes all the renormalized parameters
of  the  model,  and,  according  to  our  definitions,  the  one-loop
displacement operator is given by
\begin{eqnarray} 
D^{(1)}&=&\frac12Z^{(1)}_Hv_{\rm R}\frac\partial{\partial
v}+ Z^{(1)}_\lambda\lambda_{\rm R}\frac\partial{\partial\lambda_{\rm R}} +
Z^{(1)}_{m^2} m^2_{\rm R}\frac\partial{\partial m^2_{\rm R}} + Z^{(1)}_g
g_{\rm R}\frac\partial{\partial g_{\rm R}} + Z^{(1)}_\xi\xi_{\rm
R}\frac\partial{\partial\xi_{\rm R}} 
+ Z^{(1)}_\sigma \sigma_{\rm R}\frac\partial{\partial\sigma_{\rm R}}
\nonumber \\ 
&+& \sumi Z^{(1)}_{h_i}h_i^{\rm R}\frac\partial{\partial h_i^{\rm
R}}+\frac n2Z^{(1)}_H\; , 
\end{eqnarray} 
with  $n$ the number  of external  Higgs legs.  We emphasize  that the
above equations are not independent from each other, since, due to the
HLET, they  are related by successive differentiation  with respect to
$v$.  Notice also  that, on the RHS of  (\ref{renor}), all overlapping
divergences  should cancel, a  fact which  furnishes a  very stringent
check of  the entire  calculation.  From the  above equations  one can
infer the gauge-dependence  of the Higgs wave-function renormalization
constant at  two loops. Specifically,  expanding the last  equation of
(\ref{renor}), we find
\begin{equation} 
2Z^{(2)}_H+Z^{(2)}_\lambda =
-(Z^{(1)}_H)^2-2Z^{(1)}_HZ^{(1)}_\lambda-\frac i{6\lambda}
\left.\left\{\overline{D^{(1)}\Gamma^{(1)}_{H^4}(x_{\rm R};\mu,\epsilon)}+
\overline{\Gamma^{(2)}_{H^4}(x_{\rm
R};\mu,\epsilon)}\right\}\right|_{\sigma_{\rm R}=1}\; .  
\end{equation}
Differentiation with respect to $\xi$, taking into account that, 
in the $\overline{\rm MS}$ scheme that we are using, 
$Z_\lambda$ is GFP independent, yields the following identity 
\begin{equation}
-\frac\partial{\partial\xi} Z^{(2)}_H = (Z^{(1)}_H+Z^{(1)}_\lambda)
\frac\partial{\partial\xi}Z^{(1)}_H+\frac
i{12\lambda}\frac\partial{\partial\xi}
\left.\left\{\overline{D^{(1)}{\Gamma}^{(1)}_{H^4}(x_{\rm R};\mu,\epsilon)}+
\overline{\Gamma^{(2)}_{H^4}(x_{\rm
R};\mu,\epsilon)}\right\}\right|_{\sigma_{\rm R}=1}\; . 
\label{xifin}
\end{equation}
One  may easily  verify that  similar equations  can also  be obtained
starting from any of the first three equations of (\ref{renor}).

\begin{figure}[t]
\bce
\includegraphics[width=5.5cm]{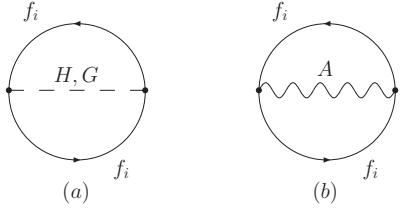}
\caption{\label{ep2loop} The two loop fermionic contribution to the
effective potential. Due to the HLET, differentiation with respect to
the Higges VEV $v$ provides the one-, two-, three- and four-Higgs
Green functions (at zero momentum).}  \ece
\end{figure}

Due  to the  HLET, from  the  diagrams contributing  to the  effective
potential we  can extract information  about the Higgs  tadpole, mass,
tri-  and  quadri-linear  couplings,  by simply  differentiating  with
respect  to  $v$.  Moreover,  since  we are  interested  only  in  the
contributions which  depend on the  Yukawa couplings, we only  need to
consider  the two-loop  fermionic  effective potential  contributions,
shown in Fig.~\ref{ep2loop}. Introducing the integrals
\begin{eqnarray}
K_{ijk}&=&\mu^{2\epsilon}\int\!\frac{d^dk}{(2\pi)^d}
\int\!\frac{d^d\ell}{(2\pi)^d}\frac{1}{(k^2-m^2_i)(\ell^2-m^2_j)
[(k+\ell)^2-m^2_k]}\nonumber \\
&=&-\frac1{(4\pi)^4}\sum_{n=i,j,k}m^2_n\left[\frac2{\epsilon^2} -
\frac2{\epsilon}\left(  
\ln\frac{m^2_n}{\bar\mu^2}-\frac32\right)\right]+\dots, \nonumber \\
L_{ij}&=&\mu^{2\epsilon}\int\!\frac{d^dk}{(2\pi)^d}
\int\!\frac{d^d\ell}{(2\pi)^d}\frac{1}{(k^2-m^2_i)(\ell^2-m^2_j)}\nonumber \\
&=&-\frac1{(4\pi)^4}m_i^2m_j^2\left[\frac4{\epsilon^2}+\frac2{\epsilon}
\left(2-\ln\frac{m^2_i}{\bar\mu^2}-\ln\frac{m^2_j}{\bar\mu^2}\right)
\right]+\dots,
\label{integrals}
\end{eqnarray}
where $\ln\bar\mu^2 = \ln(4\pi\mu^2)-\gamma_E$,  and the dots  stand for
finite parts, one finds
\begin{eqnarray}
(a)|_H&=&-\frac i2h_i^2d\Bigg[
L_{Hf_i}-\frac12L_{f_if_i}+\Bigg(2m^2_i-\frac12m^2_H\Bigg)
K_{Hf_if_i}
\Bigg],\nonumber \\
(a)|_G&=&-\frac i2h_i^2d\Bigg[
L_{Gf_i}-\frac12L_{f_if_i}-\frac12m^2_G
K_{Gf_if_i}
\Bigg],\nonumber \\
(b)&=&-i\frac{g^2}4d(d-2)\Bigg[
L_{Af_i}-\frac12L_{f_if_i}+\Bigg(m^2_i-\frac12m^2_A\Bigg)
K_{Af_if_i}\Bigg]\nonumber \\ &&+\frac
i4h_i^2d\left[m^2_AK_{af_if_i}-m^2_cK_{cf_if_i}\right] . 
\label{fermionic}
\end{eqnarray} 
The contributions  to $\overline{\Gamma^{(2)}_{H^4}}$ which  depend on
the Yukawa couplings are  simply obtained by differentiating the above
expressions  four times  with  respect to  $v$.   As far  as the  term
$\overline{D^{(1)}\Gamma^{(1)}_{H^4}}$ is  concerned, one can  use the
results  of  Appendix  A  for  the one-loop  effective  potential  and
renormalization constants (notice that  in this case, one cannot limit
one's  attention  to the  fermionic  contributions  only,  due to  the
dependence on the Yukawa couplings of the renormalization constants).

The combination  of these two  terms leads to a  massive cancellation,
yielding finally
\begin{equation}
\frac\partial{\partial\xi}Z^{(2)}_{H,h_i}=
- \frac{2g^2}{(4\pi)^4\epsilon} \sumi h^2_i,
\label{anaxi}
\end{equation}
thus establishing the GFP-dependence of $Z^{(2)}_{H,h_i}$ in the 
$\overline{R_{\xi}}$ gauges.

\begin{figure}[t]
\includegraphics[width=16cm]{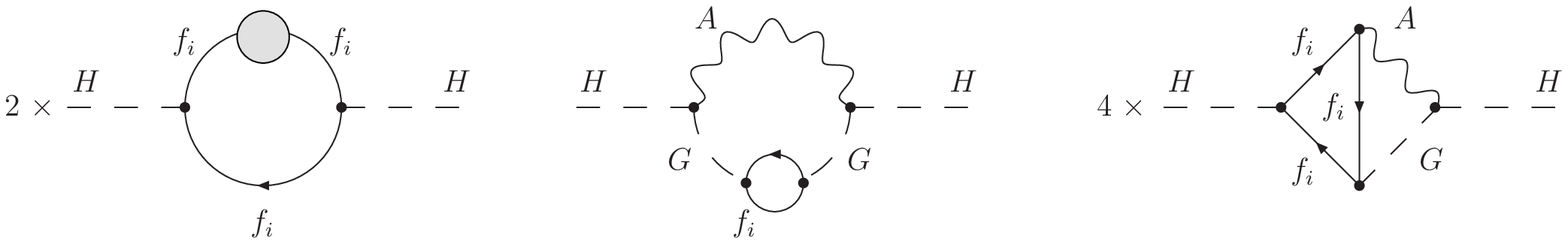}
\caption{\label{fermions} Fermion diagrams contributing
$g^2\sum_ih_i^2$ terms to the two-loop Higgs wave-function
renormalization constant $Z^{(2)}_H$ in the symmetric phase.}
\end{figure}

As a check  for the consistency of the procedure,  we can evaluate the
full two-loop Higgs wave-function renormalization constant $Z^{(2)}_H$
of our model,  through the direct calculation of  the relevant Feynman
diagrams of the Higgs-boson self-energy, shown in Fig.~\ref{fermions}.
As mentioned earlier,  we will calculate in the  symmetric phase, $v =
0$, keeping  only contributions  proportional to the  Yukawa couplings
$h_i$
 
We will  use directly the  $D$-formalism to explicitly check  that all
overlapping divergences cancel, and to get the fermionic contributions
to $Z^{(2)}_H$ through the formula
\begin{equation}
Z^{(2)}_H=\frac i2 \frac\partial{\partial p^2}
\left.\left\{\overline{D^{(1)}
\Gamma^{(1)}_{H^2}(x_{\rm
R};\mu,\epsilon)}+\overline{\Gamma^{(2)}_{H^2}(x_{\rm
R};\mu,\epsilon)}\right\}\right|_{\sigma_{\rm R}=1,\, p^2=0}. 
\end{equation}
Once again one  should keep in mind that  $\Gamma^{(1)}_{H^2}$ must be
calculated  at a  general  value  of $\sigma$  (which  means that  the
($\sigma-1$) part of the $A_\mu GH$ vertex will also contribute).  The
final result is given by
\begin{equation}
\mu^\epsilon Z^{(2)}_{H,h_i} =\frac{
- 2 g^2 (\xi + 4) \sum_ih_i^2 + 8
\sum_ih_i^4}{(4\pi)^4\epsilon}-\frac{8\sum_ih_i^4}
{(4\pi)^4\epsilon^2}\ , 
\end{equation}
which coincides with  (\ref{anaxi}) after differentiating with respect
to  $\xi$.  From  this  expression  one  can  determine  the  two-loop
anomalous   dimension  of  the   Higgs  field,   which  is   given  by
\cite{Machacek:1983tz}
\begin{eqnarray}
\gamma^{(2)}_H &=& -\frac12 \sum_n\rho_nx_n\frac\partial{\partial x_n}C_H^1,
\end{eqnarray}
where $x_n$  denotes collectively the  free parameters of  the theory,
$\rho_n\epsilon$ their mass-dimension,  and $C^1_H$ the coefficient of
the simple pole of the Higgs 1PI self-energy. Therefore
\begin{eqnarray}
\gamma^{(2)}_{H,h_i} &=& -\frac12\left(\lambda\frac\partial{\partial \lambda}
+\frac12g\frac\partial{\partial g}+\frac12\sum_i
h_i\frac\partial{\partial h_i}\right)C^1_{H,h_i} \nonumber \\ 
&=& \frac{2g^2(\xi+4)\sum_ih_i^2 -8 \sum_ih_i^4}{(4\pi)^4}\ .
\end{eqnarray}  
Evidently,  despite the  fact  that $\Delta  v=0$  {\it exactly},  the
two-loop running of $\tan\beta$ turns out to be GFP-dependent.

\section{Conclusions}

We have developed a  new formalism for determining the renormalization
and  the   GFP-dependence  of  Green   functions  to  all   orders  in
perturbation theory.   The formalism  makes use of  the fact  that the
renormalized  Green  functions are  obtained  by  displacing both  the
unrenormalized fields  and fundamental  parameters of the  theory with
respect  to the  renormalized ones.   Due  to this  property, we  have
called it the {\em displacement  operator formalism} or, in short, the
$D$-formalism.  With the help of  this formalism the CTs necessary for
the  renormalization  of  any  Green  function  can  be  unambiguously
determined, to any given  order of perturbation theory. In particular,
if  one applies  the  $D$-operator before  integrating  over the  loop
momenta, one can  systematically generate all the CTs  that would have
been  obtained   in  the  conventional   diagrammatic  framework.   We
explicitly  demonstrate the  full  potential of  the $D$-formalism  by
considering several  known examples of renormalization  of theories up
to 2-loops, such as a $\phi^4$-theory, QED, and QCD in the BFM $R_\xi$
gauge.

One of  the great advantages  of the $D$-formalism  is that it  can be
used to calculate the {\em precise form} of deformation of symmetries,
which are modified in the  process of renormalization, such as the NI.
Hence, the dependence of the {\em renormalized} Green functions on the
{\em renormalized}  GFPs can be  computed exactly, thereby  offering a
new method for evaluating the  GFP-dependence within a given scheme of
renormalization,  {\it e.g.},~the $\overline{\rm  MS}$ scheme,  the OS
scheme,  etc.  Given that  a concrete,  closed formula  describing the
deformation of the NI to all  orders exists now, one should be able to
conclusively    settle    various    formal    issues    related    to
gauge-invariance. Most notably, it would be interesting to revisit the
important question  of the all-order  gauge-invariance of the  pole of
the  unstable particles,  together with  other topics  related  to the
gauge-invariant   formulation   of   resonant  transition   amplitudes
\cite{Papavassiliou:1995fq}.

In theories  with SSB, in  addition to the  Slavnov--Taylor identities
and the  NI, the HLET plays  an important role as  well.  The ordinary
$R_\xi$  gauge  violates  the   HLET  for  off-shell  1PI  correlation
functions. In order to explore  the constraints imposed by the HLET we
have resorted to a toy field theory with SSB, the Abelian Higgs model,
which  was  quantized using  a  HLET-preserving  gauge.  An  important
consequence  of these  gauges is  that the  VEV $v$  of a  Higgs field
renormalizes multiplicatively  by the Higgs wavefunction,  so there is
no additional UV infinite shift to $v$, {\it i.e.}~$\Delta v_{\rm div}
= 0$.  Employing the $D$-formalism in the context of a HLET-preserving
gauge,  we  have  shown  that the  fundamental  quantity  $\tan\beta$,
defined  in  the two-Higgs  models,  is  GFP-dependent  at two  loops,
exactly as happens  for the usual $R_\xi$ gauge,  in which there exist
additional  UV  infinite  shifts  to  the Higgs  VEVs.   The  analysis
presented here  strongly suggests that the $\tan\beta$  cannot be made
gauge-independent   within  the   framework   of  conventional   Green
functions,  even if  the gauge-fixing  employed respects  all relevant
symmetries and  constraints, such as the HLET.  These results motivate
one  to explore  the  possibility of  defining  $\tan\beta$ at  higher
orders   through  the   GFP-independent   effective  Green   functions
constructed  by means of  the pinch  technique \cite{Cornwall:1981zr}.
In  particular, it  would be  interesting  to extend  the concept  and
construction      of     the     Higgs-boson      effective     charge
\cite{Papavassiliou:1997fn}  to the  case of  multi-Higgs  models, and
more specifically to  supersymmetric theories.  We hope to  be able to
report progress on this subjects in the near future.

The formulation developed in  this article presents novel perspectives
for  the  study  of  several  other known  topics.  Specifically,  the
$D$-formalism   may  be   used  to   systematically   investigate  the
renormalization-scheme  dependence of  1PI correlation  functions.  It
may also be employed to algebraically determine the restoring terms of
a ``bad'' UV  regularizing scheme, {\it i.e.}, a  scheme that does not
preserve the Slavnov--Taylor  identities.  Since it provides all-order
information on the renormalization  of Green functions under study, it
might  be useful  in controlling  the calculation  of non-perturbative
effects,  such  as  those  related  to  the  dynamics  of  renormalons
\cite{Lautrup:1977hs}.

The $D$-formalism can be straightforwardly extended to systematize the
procedure  of  renormalizing  non-renormalizable field  theories.   In
particular,  it may be  used to  organize the  infinite series  of CTs
needed  to  renormalize  such  theories.   But even  in  the  case  of
renormalizable  perturbative field theories  the $D$-formalism  can be
automated, for  example with  the aid of  a computational  package, to
reliably compute all  the CTs required for the  renormalization of 1PI
correlation  functions   at  high   orders.  It  would   therefore  be
interesting to explore these new horizons opening up, embarking into a
study of the onset of non-perturbative dynamics at very high orders of
perturbation theory \cite{LeGuillou:1990nq}.

\acknowledgments We  thank Boris Kastening  for useful communications.
The work of J.P. has  been supported by the Grant CICYT FPA2002-00612,
and the  work of AP is supported  in part by the  PPARC research grant
PPA/G/O/2000/00461.   D.B.  thanks   the  Physics  Department  of  the
University of Manchester and  the Departamento de F\`\i sica Te\`orica
of  the University  of  Valencia, where  part  of this  work has  been
carried  out,   for  the  warm  hospitality   and  financial  support.
A.P. thanks the Departamento de F\`\i sica Te\`orica of the University
of Valencia for the kind  hospitality extended to him during the early
stages   of   this   work.    All  diagrams   drawn   using   JaxoDraw
\cite{Binosi:2003yf}.

\newpage
\begin{appendix}

\section{Feynman rules and renormalization of the Abelian Higgs model}

\begin{figure}[b]
\includegraphics[width=16cm]{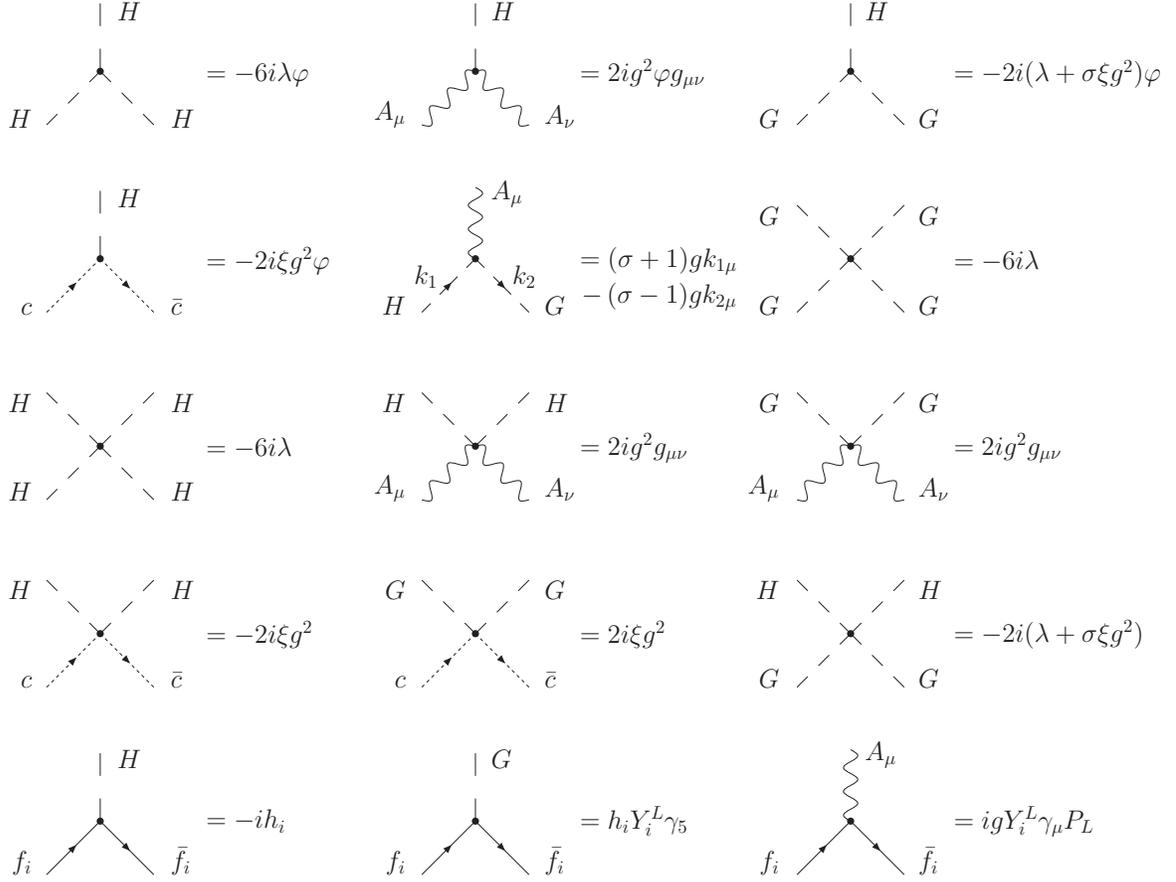}
\caption{\label{feynmanrules} The Feynman rules for the Abelian Higgs
model in the HLET type of gauge.}
\end{figure}

The Feynman rules derived from the Lagrangian density given in 
(\ref{LHLET}) are listed in Fig.~\ref{feynmanrules}.
The model can be renormalized with the renormalization condition
$\sigma_{\mathrm R}=1$, by introducing the following renormalization
constants
\begin{eqnarray}
& & H=Z_H^\frac{1}{2}H_{\mathrm R}, \qquad
G=Z_G^\frac{1}{2}G_{\mathrm R}, \qquad
A^\mu=Z_A^\frac{1}{2}A^\mu_{\mathrm R}, \qquad
\bar c c=Z_c\bar c_{\mathrm R} c_{\mathrm R}, \qquad
v=Z_H^\frac{1}{2} v_{\mathrm R}, \nonumber \\ 
& & m^2=Z_{m^2}m^2_{\mathrm R},\qquad
\lambda=Z_\lambda\lambda_{\mathrm R},\qquad\quad
g=Z_gg_{\mathrm R},\qquad\quad
\xi=Z_\xi\xi_{\mathrm R},\qquad\quad
\sigma=Z_\sigma\sigma_{\mathrm R}, \nonumber \\
& & f^{L}_i=(Z_i^L)^{\frac12}f^L_{i{\rm R}}, \qquad 
f^{R}_i=(Z_i^R)^{\frac12}f^R_{i{\rm R}}\qquad
h_i=Z_{h_i}h^{\mathrm R}_i\; . 
\end{eqnarray}
Notice that  left and right  fermions get renormalized  with different
renormalization constants.

The   determination  of   the  renormalization   constant   above,  is
simplified, due to the HLET: for example, the knowledge of the fermion
self-energy   will   automatically   imply   the  knowledge   of   the
Higgs-fermion-fermion vertex through the differentiation of the former
with respect  to the  Higgs vev $v$  (notice that, in  this particular
case, we  would find immediately  that this vertex is  one-loop finite
within the HLET gauges).

In table \ref{t1}, we report  all the divergent parts for the one-loop
Green functions of the model.
\begin{table}[t]
\begin{tabular}{c|l||c|l}
\footnotesize{$\overline{\Gamma}$} &
\multicolumn{3}{l}{\footnotesize{
$\frac i{(4\pi)^2\epsilon}\frac12\left\{\left(
10\lambda^2+2\xi g^2\lambda+3g^4-4\sumi h_i^4\right) v^4
+\left(8\lambda+2\xi g^2\right)m^2 v^2+2m^4\right\}$}}\\ \hline
\footnotesize{$\overline{\Gamma_H}$} &
\multicolumn{3}{l}{\footnotesize{
$\frac {2i}{(4\pi)^2\epsilon}\left\{\left(
10\lambda^2+2\xi g^2\lambda+3g^4-4\sumi h_i^4\right) v^3
+\left(4\lambda+\xi g^2\right)m^2 v\right\}
$}} \\ \hline\hline
\footnotesize{$\overline{\Gamma_{H^2}}$} &
\multicolumn{3}{l}{\footnotesize{$\frac
{-i}{(4\pi)^2\epsilon}\left\{\left[(6+2\xi)g^2-4\sumi
h_i^2\right]q^2\right.$}}\\ 
& \multicolumn{3}{l}{\footnotesize{$\left.-(8\lambda+2\xi
g^2)m^2-\left(60\lambda^2+12\lambda\xi g^2+18g^4-24\sumi h_i^4\right)
v^2\right\}$}} \\ \hline 
\footnotesize{$\overline{\Gamma_{G^2}}$} &
\multicolumn{3}{l}{\footnotesize{
$\frac i{(4\pi)^2\epsilon}\left\{\left[6(\xi-1)g^2+
4\sumi h_i^2\right]q^2\right.$}}\\
& \multicolumn{3}{l}{\footnotesize{
$+\left.(8\lambda-6\xi g^2)m^2+\left(20\lambda^2+4\lambda\xi
g^2+6g^4+6\xi^2g^4-8\sumi h_i^4\right) 
 v^2\right\}$}} \\ \hline
\footnotesize{$\overline{\Gamma_{A_\mu A_\nu}}$} &
\multicolumn{3}{l}{\footnotesize{
$\frac i{(4\pi)^2\epsilon}\left\{\left[ 
-\frac{10}3g^2q^2-2\left(3g^2+\xi g^2-2\sumi h_i^2\right)g^2
v^2\right]g_{\mu\nu} 
+\frac{16}3g^2 q_\mu q_\nu\right\}$}}\\ \hline
\footnotesize{$\overline{\Gamma_{A_\mu G}}$} &
\multicolumn{3}{l}{\footnotesize{
$\frac{2}{(4\pi)^2\epsilon}\left\{
2\lambda+3\xi g^2-3g^2+2\sumi h_i^2\right\}g v q_\mu$}} \\ \hline
\footnotesize{$\overline{\Gamma_{c\bar c}}$} &
\multicolumn{3}{l}{\footnotesize{$\frac
i{(4\pi)^2\epsilon}\left\{4\lambda\xi g^2 
+6\xi^2 g^4\right\} v^2$}} \\ \hline
\footnotesize{$\overline{\Gamma_{f_i\bar f_i}}$} &
\multicolumn{3}{l}{\footnotesize{
$\frac i{(4\pi)^2\epsilon}\left\{\qslush P_L(2g^2\xi+2h_i^2)
+\qslush P_R(2h_i^2)\right\}$}} \\ \hline\hline
\footnotesize{$\overline{\Gamma_{H^3}}$} &
\footnotesize{$\frac{12i}{(4\pi)^2\epsilon}\left\{
10\lambda^2+2\lambda\xi g^2+3g^4-4\sumi h_i^4
\right\} v$} & \footnotesize{$\overline{\Gamma_{HA_\mu A_\nu}}$} &
\footnotesize{$\frac{-4i}{(4\pi)^2\epsilon}\left\{
3g^2+\xi g^2-2\sumi h_i^2\right\}g^2 v g_{\mu\nu}$}\\ \hline
\footnotesize{$\overline{\Gamma_{H^4}}$} &
\footnotesize{$\frac{12i}{(4\pi)^2\epsilon}\left\{
10\lambda^2+2\lambda\xi g^2+3g^4-4\sumi h_i^4
\right\}$} & \footnotesize{$\overline{\Gamma_{H^2A_\mu A_\nu}}$} & 
\footnotesize{$\frac{-4i}{(4\pi)^2\epsilon}\left\{
3g^2+\xi g^2-2\sumi h_i^2\right\}g^2g_{\mu\nu}$}\\ \hline
\footnotesize{$\overline{\Gamma_{G^4}}$} &
\footnotesize{$\frac{12i}{(4\pi)^2\epsilon}\left\{10\lambda^2-6\lambda\xi g^2
+3g^4-4\sumi h_i^4\right\}$} &
\footnotesize{$\overline{\Gamma_{G^2A_\mu A_\nu}}$} &  
\footnotesize{$\frac{-4i}{(4\pi)^2\epsilon}\left\{3\left(1-\xi\right)g^2-2\sumi
h_i^2\right\} g^2g_{\mu\nu}$}\\ \hline 
\footnotesize{$\overline{\Gamma_{HG^2}}$} &
\footnotesize{$\frac{4i}{(4\pi)^2\epsilon}\left\{
10\lambda^2+2\lambda\xi g^2+3g^4\right.$} &
\footnotesize{$\overline{\Gamma_{Hc\bar c}}$} &  
\footnotesize{$\frac{4i}{(4\pi)^2\epsilon}\left\{2\lambda\xi g^2
+3\xi^2 g^4\right\}v$} \\
& \footnotesize{$\left.+3\xi^2g^4-4\sumi h_i^4
\right\}v$} & & \\ \hline
\footnotesize{$\overline{\Gamma_{H^2G^2}}$} &
\footnotesize{$\frac{4i}{(4\pi)^2\epsilon}\left\{
10\lambda^2+2\lambda\xi g^2+3g^4+3\xi^2g^4-4\sumi h_i^4
\right\}$} & \footnotesize{$\overline{\Gamma_{H^2c\bar c}}$} & 
\footnotesize{$\frac{4i}{(4\pi)^2\epsilon}\left\{2\lambda\xi g^2
+3\xi^2 g^4\right\}$} \\ \hline
\footnotesize{$\overline{\Gamma_{A_\mu HG}}$} &
\footnotesize{
$\frac{-2g}{(4\pi)^2\epsilon}\left\{\left(
2\lambda+\xi g^2+3g^2-2\sumi h_i^2\right)k_{1\mu}\right.$} &
\footnotesize{$\overline{\Gamma_{G^2c\bar c}}$} &  
\footnotesize{$\frac{-4i}{(4\pi)^2\epsilon}\left\{2\lambda\xi g^2
-\xi^2 g^4\right\}$} \\ 
& \footnotesize{
$\left.-\left(
2\lambda+3\xi g^2-3g^2+2\sumi h_i^2\right)k_{2\mu}\right\}$} & & \\ \hline
\end{tabular}
\caption{\label{t1}The one-loop divergent parts of the Green functions
of the Abelian Higgs model in the HLET preserving gauge (with
$\sigma=1$).}
\end{table}
Using these results, one finds  
\begin{eqnarray}
\mu^\epsilon Z_H &=& 1 + 
\frac{(6+2\xi)g^2-4\sumi h_i^2}{(4\pi)^2\epsilon}\ , \nonumber\\ 
\mu^\epsilon Z_G &=& 
1+\frac{(6-6\xi)g^2-4\sumi h_i^2}{(4\pi)^2\epsilon}\ , \nonumber\\
\mu^\epsilon Z_A &=& 1+\frac{-\frac{10}3g^2}{(4\pi)^2\epsilon}\ , \nonumber\\ 
\mu^\epsilon Z_c &=& 1\; , \nonumber\\ 
\mu^\epsilon Z_i^L &=& 1+\frac{-2g^2\xi-2h_i^2}{(4\pi)^2\epsilon}\ ,\nonumber\\
\mu^\epsilon Z_i^R &=& 1+\frac{-2h_i^2}{(4\pi)^2\epsilon}\ ,\nonumber\\ 
Z_{m^2} &=& 1+\frac{(8\lambda-6g^2+4\sumi h_i^2)}{(4\pi)^2\epsilon}\ , 
             \nonumber\\ 
\mu^{-\epsilon} Z_\lambda &=& 1+\frac{20\lambda-12g^2 + 
8\sumi h_i^2+\frac6\lambda g^4 
-\frac8\lambda\sumi h_i^4}{(4\pi)^2\epsilon}\ , \nonumber\\
\mu^{-\frac\epsilon2} Z_g &=& 
1+\frac{\frac{5}3g^2}{(4\pi)^2\epsilon}\ ,\nonumber\\ 
\mu^{-\frac\epsilon2} Z_{h_i} &=& 1 + 
\frac{-3g^2+2h_i^2+2\sum_{j=1}^2h_j^2}{(4\pi)^2\epsilon}\ ,\nonumber\\ 
Z_\xi&=&1+\frac{4\lambda + 
4\xi g^2-\frac{28}3g^2+4\sumi h_i^2}{(4\pi)^2\epsilon}\ , \nonumber\\
Z_\sigma &=& 1+\frac{4\lambda+(-6+6\xi)g^2+4\sumi
h_i^2}{(4\pi)^2\epsilon}\ .
\end{eqnarray}
Notice  that  $Z_A^{\frac12}Z_g=1$  as  it  should  due  to  the  Ward
identities of the theory.  Combining  the divergent parts of the Green
functions   given  in   table  \ref{t1},   together  with   the  above
renormalization constants,  one can  explicitly check the  validity of
the renormalized NI of Eq.(\ref{1ldefNI}) at one loop.

The effective potential  is particularly useful in the  context of the
HLET-preserving gauges, furnishing a substantial amount of information
with   relatively   moderate   effort   \cite{Kastening:1993zn}.    In
dimensional regularization, the one-loop effective potential reads
\begin{eqnarray}
V^{(1)}&=&\frac
i{4(4\pi^2)}\Bigg[ {m^4_H}\left(\frac32-\ln\frac{m^2_H}{\bar\mu^2}\right)+
{m^4_G}\left(\frac32-\ln\frac{m^2_G}{\bar\mu^2}\right)+{m^4_A}
\left(\frac52-3\ln \frac{m^2_A}{\bar\mu^2}\right)\nonumber\\[2mm] 
&& +\,  \frac{m^4_c}{\sigma^2} \left( \frac32 -
\ln\frac{m^2_c}{\sigma^2\bar\mu^2}\right) 
-2m^2_c\left(\frac32-\ln\frac{m^2_c}{\bar\mu^2}\right)-4\sumi
m^4_i\left(1-\ln\frac{m^2_i}{\bar\mu^2}\right)\Bigg]\nonumber\\ 
&& +\:
\frac{i}{2(4\pi)^2\epsilon}\Bigg( m^4_H+m^4_G+3m^4_A+\frac{m_c^4}{\sigma^2}
-2m_c^4-\sumi m^4_i \Bigg)\;,
\end{eqnarray}
where $\ln\bar\mu^2=\ln(4\pi\mu^2)-\gamma_E$. As  stated earlier it is
important to  keep explicitly the  dependence on $\sigma$,  since this
will play a crucial role  in cancelling the overlapping divergences in
our  two-loop expressions,  when  the $D$-formalism  is applied  after
integration.

\begin{figure}[t]
\begin{center}
\includegraphics[width=5cm]{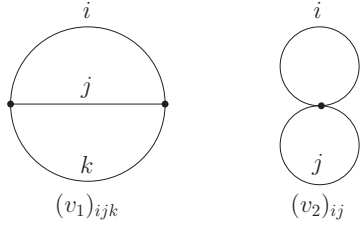}
\caption{\label{eptop} Topologies contributing to the two loop
effective potential. $i$, $j$, and $k$ labels all the possible field
combinations allowed by the Feynman rules.}
\end{center}
\end{figure}

At  two loops,  there are  two  basic topologies  contributing to  the
effective potential, shown in Fig.~\ref{eptop}.  It turns out that all
these two  loop diagrams  can be expressed  in terms of  the integrals
$L_{ij}$ and $K_{ijk}$ introduced in Eq.(\ref{integrals}).  In table 2
we report the results of the (scalar and vector) diagrams.
\begin{table}[b]
\begin{tabular}{c|l||c|l}
\footnotesize{$(v_1)_{H^3}$} &
\multicolumn{3}{l}{\footnotesize{$3iv^2\lambda^2K_{HHH}$}} \\ \hline 
\footnotesize{$(v_1)_{HG^2}$} &
\multicolumn{3}{l}{\footnotesize{$iv^2(\lambda+\xi g^2)^2K_{HGG}$}} \\
\hline 
\footnotesize{$(v_1)_{Hc\bar c}$} &
\multicolumn{3}{l}{\footnotesize{$-2iv^2\xi^2 g^4K_{Hc\bar c}$}} \\
\hline 
\footnotesize{$(v_1)_{HAG}$} & \multicolumn{3}{l}{\footnotesize{$2ig^2\left\{
m^2_HK_{HAG}+L_{AG}+\frac12\xi L_{Gc}-\frac12
L_{GA}-\frac14(1-\xi)\left[L_{cG}+L_{HG}- 
L_{cH}\right.\right.$}}\\
& \multicolumn{3}{l}{\footnotesize{$\left.(2m^2_G-2m^2_H-m^2_c-m^2_A)
K_{HGc}\right]+\frac{m^2_G-m^2_H-m^2_A}{4m^2_A}\left[L_{GA}+ 
L_{Hc}-L_{HA}-L_{Gc}\right]$}}\\
& \multicolumn{3}{l}{\footnotesize{$-\frac{(m^2_G-m^2_H-m^2_A)^2}{4m^2_A}\left[
K_{HGA}-K_{HGc}\right]\big\}$}}\\ \hline 
\footnotesize{$(v_1)_{HA^2}$} & \multicolumn{3}{l}{\footnotesize{$iv^2g^4\big\{
(d-2)K_{HAA}+2\xi K_{HAc}-
\frac1{2m^2_A}(1-\xi)[L_{cA}+L_{Hc}-L_{cc}-L_{HA}] + 
\frac12(1-\xi)^2K_{Hcc}$}}\\ 
& \multicolumn{3}{l}{\footnotesize{$
+ \frac{m^2_H-2m^2_A}{4m^4_A}[L_{AA}+L_{cc}-2L_{Ac}] +
\frac1{2m^2_A}(1-\xi)(3m^2_A-2m^2_H+m^2_c)
\left[K_{HAc}-K_{Hcc}\right]$}}\\ 
& \multicolumn{3}{l}{\footnotesize{$+\frac{(m^2_H-2m^2_A)^2}{4m^4_A}\left[
K_{HAA}+K_{Hcc}-2K_{HAc}\right]\big\} $}}\\ \hline\hline
\footnotesize{$(v_2)_{H^2}$} & \footnotesize{$i\frac{3\lambda}4
L_{HH}$\hspace{4.5cm}} & \footnotesize{$(v_2)_{G^2}$} &
\footnotesize{$i\frac{3\lambda}4 L_{GG}$} \\ \hline 
\footnotesize{$(v_2)_{HG}$} & \footnotesize{$i\frac{\lambda+\xi g^2}2
L_{HG}$} & \footnotesize{$(v_2)_{GA}$} &
\footnotesize{$i\frac{g^2}2\big\{(d-1)L_{GA}+\xi L_{Gc}\big\}$} \\
\hline 
\footnotesize{$(v_2)_{Hc}$} & \footnotesize{$-i\xi g^2L_{Hc}$} &
\footnotesize{$(v_2)_{Gc}$} & \footnotesize{$i\xi g^2L_{Gc}$} \\
\hline 
\end{tabular}
\caption{\label{t2} Scalar and vector two loop contributions to the
effective potential (with $\sigma=1$). The fermionic contributions are
given in Eq.(\ref{fermionic}) of the text.}
\end{table}
Using  these results  in  conjunction with  Eq.(\ref{xifin}), one  can
obtain the  full gauge-dependence of the  two-loop Higgs wave-function
renormalization and anomalous dimension. Specifically,
\begin{eqnarray}
\frac\partial{\partial\xi}Z^{(2)}_H&=&
\frac{-2g^2\sumi h^2_i-4\lambda g^2-6\xi g^4+2g^4}{(4\pi)^4\epsilon}+ 
\frac{8\lambda g^2+24\xi g^4}{(4\pi)^4\epsilon^2}, \nonumber \\
\frac\partial{\partial\xi}\gamma^{(2)}_H&=&\frac{2g^2\sumi
h^2_i+4\lambda g^2+6\xi g^4-2g^4}{(4\pi)^4\epsilon}\; .
\end{eqnarray}
We  end by  reporting for  completeness the  full two-loop  Higgs wave
function renormalization and anomalous dimension:
\begin{eqnarray}
\mu^\epsilon Z^{(2)}_H &=&\frac{-4\lambda-4\lambda\xi g^2+(-2+2\xi-3\xi^2)g^4-
2\xi g^2\sum_ih_i^2 -8 g^2\sum_ih_i^2 + 8
\sum_ih_i^4}{(4\pi)^4\epsilon}\nonumber \\ 
&+&\frac{8\lambda\xi g^2+(12+12\xi^2)g^4-12g^2\sum_ih_i^2
-8\sum_ih_i^4} {(4\pi)^4\epsilon^2}, \nonumber \\ 
\gamma^{(2)}_H &=& \frac{2\lambda  +(3\xi^2-2\xi+2)g^4+4\lambda\xi
g^2+2g^2(\xi+4)\sum_ih_i^2 -8 \sum_ih_i^4}{(4\pi)^4}\; . 
\end{eqnarray} 

\end{appendix}


\begin{thebibliography}{99}
\bibitem{footnote1}
Although  it  is easy to
incorporate in our formalism, we will not  consider for simplicity the
cosmological    constant,    which  only  becomes    relevant   to the
renormalization of vacuum energy graphs.
\bibitem{footnote2}
If  there  is mixing  between  different types  of fields
$\phi_i$, the  first term should  be replaced by  $\sum_{j,i}\, \delta
\phi_{ji} \partial/\partial \phi_{i{\rm R}}$, with $\delta \phi_{ij} =
[(Z^{\frac12}_\phi)_{ij} - \delta_{ij}]\,\phi_{i{\rm R}}$.
\bibitem{footnote3}
Up to two loops, this is not such a serious complication, 
since one needs the explicit $\xi$ dependence of one-loop diagrams only.
\bibitem{footnote5}
In the presence
of  mixing  of  different  fields,   the wavefunction renormalizations
$Z^{\frac12}_{\phi_i}$    are  to be    replaced  by  matrices  of the
form~$(Z^{\frac12}_\phi)_{ij}$.
\end{thebibliography}
\end{document}